# Tutorial Review of Mixing in a Rotating Soft Microchannel under Electrical Double Layer Effect: A Variational Calculus Approach

*Harshad Sanjay Gaikwad, Pranab Kumar Mondal\**

*Microfluidcs and Microscale Transport Processes Laboratory,*
*Department of Mechanical Engineering, Indian Institute of Technology Guwahati, Assam,*
*India – 781039*

---
*Corresponding author, email: mail2pranab@gmail.com, pranabm@iitg.ernet.in

**Abstract**

We study the effect of the grafted polyelectrolyte layer on the flow dynamics, and its consequences on underlying mixing in the rotating microfluidic channel. For this analysis, the method used by Sadeghi *et al.* (*J. Fluid Mech.,* vol. 887, 2020, pp. A13; *Phys. Rev. Fluids.,* vol. 4 (6), 2019, 063701-23), is modified by incorporating the non-linear effect stemming from the polyelectrolyte layer induced electrostatics to solve the coupled system of equations, integrated with the non-homogeneous boundary conditions. This method is used to obtain the velocity distribution in the asymptotic limit of geostrophic plug flow under the framework of variational calculus approach. We analyze the mixing dynamics from the perspective of both qualitative assessment and quantitative evaluation. For the qualitative estimation, we focus on the Poincaré map analysis, while the entropy of mixing approach is used for the mixing quantification. Results show that the grafted polyelectrolyte layer in contact with the ionic solution leads to the development of an electrical double layer, which upon interacting with the external electric field, strengthens the electroosmotic pumping in the fluidic channel. Such polyelectrolyte layer modulated strong electroosmotic pumping together with its intrinsic feature of offering a frictional drag to the underlying transport helps to modulate the primary as well as the secondary flows in the channel under the influence of rotational forces. With an alteration in the electroosmotic pumping and frictional drag force, tunable through the thickness of the grafted polyelectrolyte layer, we obtain three different types of secondary flow vortex configurations viz., a standard double-vortex, dumbbell-shaped vortex and the transition state between the formers. A significant change in the structure and strengths of these vortices constitutes a variation in the periodic motion of the secondary flow, which in turn, modulates the chaotic mixing in the present flow configuration. We establish that, for small Debye-Hückel parameter of electrolyte and higher polyelectrolyte layer thickness, the strong dumbbell-shaped vortices being formed enhance the chaotic nature and increase the underlying mixing performance. Whereas in case of very small polyelectrolyte layer thickness and higher frictional drag, the standard double-vortex configuration helps to improve the quality of mixing by increasing the strength of the recirculation zones.

## 1 Introduction

The rotational fluidics in micro-electro-mechanical-systems (MEMS) and CD-based lab-on-a-chip (LOC) platforms offers several advantageous features and have made significant advances in the field over the past few decades (Ducrée *et al.* 2006; Duffy *et al.* 1999; Stone *et al.* 2004). Of these features, the periodic helical motion of the secondary flow vortices set by the Coriolis force and the lateral wall confinement renders an intriguing nature to the mixing in the rotational microfluidics (Ng & Qi 2015). It may be mentioned here that the underlying flow dynamics in the miniaturized assays integrated on the rotating platforms is encountered with a few serious challenges. First, for example, the increasing magnitude of the Coriolis force adversely affects the axial momentum and triggers a rigid body type motion with increasing magnitude of the rotational speed beyond no-flow limits (Stewartson 1957). Second, ensuring a considerable amount of mixing efficiency in rotating devices/systems leads to the intervention of the geometrical and physical features of the centrifuges. Alongside above all issues, the direct exposure of fluid samples, typically bio-samples, to the binding surfaces of rotating micro-total-analysis-systems (µTAS), leads to a significant deterioration of samples' quality as well.

One of the promising solutions to overcome these challenges could be the application of a grafted soft polyelectrolyte layer[1] (PEL) on the walls of rotating microchannels. It may be mentioned in this context here that with the advent of nanotechnology, several methods have shown their potential applicability in grafting PEL of desirable thickness at the wall of microfluidic channels in a synergetic way (Das *et al.* 2015). Quite notably, such technology at the interfacial scale provides tremendous benefits in different applications like biological analysis, immunoassays, the biochemistry analysis of biofluids, to name a few (Das *et al.* 2015). The pivotal objective behind the use of grafted PEL at the walls of the narrow fluidic assays has been the exploration of the enriched physicochemical interaction of the PEL macromolecules with the electrolyte ions, and its affluent consequences on the underlying flow dynamics (Keh & Ding 2003; Ohshima & Kondo 1990). Important to mention, the imposition of soft PEL at the walls of the fluidic pathways has demonstrated distinguishable outcomes on account of the notable fluidic functionalities, attributable primarily to the complex interactions between the polymeric macromolecules and the electrolyte ions (Chanda *et al.* 2014; Chen & Das 2016, 2017; Kaushik *et al.* 2019; Patwary *et al.* 2016; Sadeghi 2018; Sadeghi *et al.* 2019, 2020). One of the significant advantages of using grafted PEL in the narrow fluidic pathways is its contributing effect towards the flow controllability. Physico-chemical structure of the layered polymeric patches together with the interaction between the polymeric macromolecules and electrolyte ions impacts on the flow dynamics non-intuitively at the microfluidic scale, leading to a greater degree of flow control in the narrow fluidic confinements (Das *et al.* 2015; Reshadi & Saidi 2019; Sadeghi *et al.* 2020). It is worth adding here that the flow dynamics in a soft PEL grafted narrow channel, embedded in the rotating platform, upon interacting with the Coriolis force stemming from the rotational effect, establishes a necessary fluidic functionality of the enhanced vortical flow[2] in the course of flow. Moreover, these soft polyelectrolyte layers, mainly composed of proteins like basic structure, furnish the smart and protective cushioning to the bio-samples from their undesirable degradation upon contact with the solid surface (Das *et al.* 2015).

---

[1] Microfluidic channels with built-in soft polymeric layers are also known as soft microchannels. Polymeric layer following the chemical interaction with the electrolyte solution forms polyelectrolyte layer or PEL in short.

Research in the paradigm of rotational fluidics has also advanced extensively to explore the effect of electric field on the underlying transport phenomena in recent years (Abhimanyu *et al.* 2016; Kaushik *et al.* 2017a, 2017b; Kaushik & Chakraborty 2017; Ng & Qi 2015). The application of an electric field on the microflows in the rotating platform offers a few distinctive transport features. Among the several advantages, notable features include the electroosmotic flow pumping (EOF pumping) of liquid following the phenomenon of electrical double layer (EDL) effect, the inevitable Joule heating effect dries out the dissolved air/gas bubbles from the fluid being transported (Chang & Wang 2011; Ng & Qi 2015; Wang *et al.* 2004). It may be mentioned in this context here that the electroosmotic pumping not only enhances the net throughput but also, upon interacting with the rotation induced forcing, promotes the onset of double-vortex structures in the fluidic pathways (Ng & Qi 2015). Notably, the electrical double layer effect strengthens the secondary flow vortices being formed in the rotating channel as well. Quite interestingly, the polyelectrolyte layers upon interacting with the electrolytic solution have shown promising potential to establish the electrical double layer effect, for instances, see Refs. (Chanda *et al.* 2014; Chen & Das 2016, 2017; Das *et al.* 2015; Patwary *et al.* 2016; Poddar *et al.* 2016)). The attributable physical reasoning behind this observation is their inherent physicochemical properties (Das *et al.*

---

[2] Vortical flow refers to the flow with the appearance of vortices.

2015). The grafted polyelectrolyte layer modulated electrostatics under the application of the electric field brings about a rich flow dynamics in the rotating fluidic environment, attributed primarily to the complex coupling of different forces that are in use. It is anticipated that the stronger electroosmotic effect, together with the frictional drag offered by the PEL, can modulate the rotational flow dynamics non-trivially, leading to an enhancement in the mixing performance following the formation of dumbbell-shaped vortices in the field. Nevertheless, no analysis has been reported to date considering the aforementioned effects in tandem in the rotational microfluidic platform.

Here we analyze the effect of grafted PEL on the flow dynamics and its consequences on the underlying mixing in a rotating rectangular microfluidic channel under the influence of electrical forcing. Our analysis primarily focuses on the low Rossby number flow and develops a theoretical framework towards modelling the electrohydrodynamics in a rotating microfluidic channel having grafted PEL at the walls. The solution to the set of governing transport equations using the boundary conditions specific to the problem under present consideration involves added complexities, and it is, therefore, necessary to pay attention to several involved issues that are in demand. These critical issues include the non-homogeneous boundary conditions imposed by the PEL electrostatics, the Darcy frictional drag at the PEL and EL(electrolyte layer) interface, and the coupled system of equations owing to the presence of generated pressure gradients in the lateral and transverse directions under the effect of lateral confinement (Ng & Qi 2015). A recent study has shown the efficacy of the variational calculus method to tackle the issues, as mentioned above, quite effectively (Sadeghi 2018; Sadeghi *et al.* 2019, 2020). From the results of our study, we also establish that the approximate solutions provided by the variational calculus method are consistent with the imposed boundary conditions at the walls and PEL-EL interface and, at times, including the effect of lateral wall confinement accurately. To the best of our knowledge, the present study, for the first time, focuses on the variational calculus method for solving the underlying electrohydrodynamics in a rotating microfluidic channel having built-in soft layers. On solving the Eulerian velocity fields, we further use the Lagrangian approach to understand the mixing in the present rotating fluidic system. In the Lagrangian approach, the Poincaré maps offer qualitative information about the mixing phenomenon, while the entropy of the mixing provides quantification of mixing performance (Phelan *et al.* 2008; Stone & Stone 2005). By establishing the accuracy of the present theoretical approach, we discuss some new physical insights obtained from this study in the forthcoming sections. We first discuss the flow dynamics thoroughly and then discuss the mixing performance of the proposed micromixing assay. We believe that the findings of the present endeavor will provide an essential basis for the efficient design of the rotating microfluidic platform towards meeting the demand of desired fluidic functionalities, including enhanced mixing without compromising the net throughput.

## 2 Electrohydrodynamics in rotating soft microchannel

### 2.1 *Description of rotating soft microchannel*

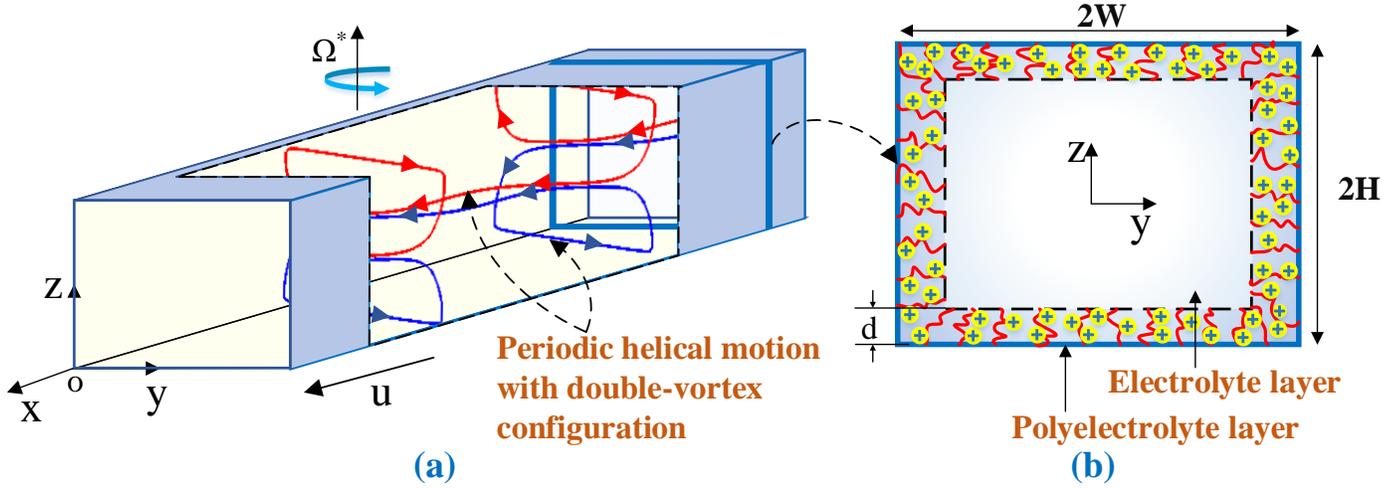

FIGURE 1 (color online). **Rotating soft microchannel:** Schematic in (a) showing the soft microchannel rotating about $z$ axis with rotational speed $\Omega^*$. The axial flow takes place in the positive $x$ direction while the transverse and vertical flows occur in $y$ and $z$ directions, respectively. The periodic helical motion adopted by the fluid particles is shown in the cut section of (a). **Channel cross-section:** Cross-section of the channel shown in (b) with two separate layers of electrolyte and polyelectrolyte in $y-z$ plane. The height and width of the channel is '$2H$' and '$2W$' respectively, whereas the thickness of PEL is '$d$'.

Figure 1 shows the schematic of the soft microfluidic channel, rotating about $z$ axis at a steady rotational speed $\mathbf{\Omega} = (0, 0, \Omega^*)$. The walls of the channel are grafted with the soft PEL of thickness $d$ (see figure 1(b)). As shown in figure 1, the electroosmotic flow stems from the electrostatic interaction between the polyelectrolyte layer modulated EDL and the externally applied electric field $\mathbf{E} = (E_x, 0, 0)$. It assists the pressure-driven flow in the duct, which has a rectangular cross-section of height '$2H$' and width '$2W$' as shown in figure 1(a). The pressure-driven flow is maintained by the axial gradient of the gyrostatic pressure $\mathbf{p}$, which is the net effect of applied axial pressure gradient and the centrifugal acceleration (Kaushik *et al.* 2017b, 2017a). The axial flow developed from these primary actuation parameters is translationally symmetric along the duct, while the secondary flow with the vortical flow structure is symmetric about the transverse centerline of the channel (Hart 1971; Johnston *et al.* 1972; Lezius & Johnston 1976).

In the rotating frame of reference, the vortical flow takes place due to the Coriolis force. If the channel is observed from the inlet, the Coriolis force sweeps the fluid towards the right wall of the channel, from where it turns back to the horizontal walls, and then to the left wall. This typical flow trajectory gives rise to a double-vortex configuration for higher rotational speed (Kheshgi & Scriven 1985; Speziale 1982), causing periodic helical motion of the fluid particles as shown in figure 1(a). The channel is considered to be sufficiently long, such that the flow structure at the intermediate cross-section ( $y$ - $z$ plane - a region of our interest in this study) remains almost unaffected by the entry and exit effects of the channel. The flow in the channel is assumed to be free from convective inertia for a very small Reynolds number $\text{Re}(\ll 1)$, typically considered in the microfluidic analysis (Ng & Qi 2015). Note that due to low $\text{Re}(\ll 1)$, the underlying transport phenomenon falls in the regime of low Rossby number flow $(\text{Ro} < 1)$ as well. We may mention here that the Rossby number is the ratio of inertia force to the Coriolis force (Ng & Qi 2015). This order analysis is suggestive of the dominating effect of the Coriolis force on the underlying transport of the problem under present consideration.

### 2.2    *PEL Electrostatics*

As mentioned, the grafted polyelectrolyte layer following the electrostatic interaction with the ionic solution leads to the development of the electrical double layer. Since the electroosmotic force contributes to the flow actuation mechanism in the present configuration, we here focus on the description of the potential distribution both in the electrolyte layer and polyelectrolyte layer by appealing to the Poisson equation (Chanda *et al.* 2014).

Electrolyte layer

$$\frac{\partial^2 \psi}{\partial y^2} + \frac{\partial^2 \psi}{\partial z^2} = -\frac{\rho_e}{\varepsilon_0 \varepsilon_r} \tag{2.1.a}$$

Polyelectrolyte layer

$$\frac{\partial^2 \psi}{\partial y^2} + \frac{\partial^2 \psi}{\partial z^2} = -\frac{\rho_e + ZeN}{\varepsilon_0 \varepsilon_r} \tag{2.1.b}$$

In equations (2.1.a) and (2.1.b), $\psi$ is the electrostatic potential, $\varepsilon_0$ is the permittivity of the free space, $\varepsilon_r$ is the relative permittivity of the electrolyte, $Z$ is the valence on PEL ions, $e$ is the electronic charge and $N$ is the number density of the PEL ions. The charge density $(\rho_e)$ of the electrolyte is given by (Masliyah & Bhattacharjee 2006; R. J. Hunter 1981),

$$\rho_e = ze(n_+ - n_-) \tag{2.2}$$

where $z$ is the valence of electrolyte and $n_\pm$ is the number density of positive/negative electrolyte ions. Under the thermodynamic equilibrium, the number density $n_\pm$ of the electrolyte ions is given by the Boltzmann charge distribution, i.e., $n_\pm = n_\infty \exp[\mp(ze\psi)/(k_B T)]$ where $n_\infty$ is the neutral charge density of the electrolyte, $k_B$ is the universal Boltzmann constant and $T$ is the absolute temperature of the fluid (Masliyah & Bhattacharjee 2006; R. J. Hunter 1981). Note that the validity leading to the applicability of the Boltzmann charge distribution in the context of the present analysis is detailed in the supporting information part of this paper (section S1). Now, by using this charge distribution given in equation (2.2), we can modify equations (2.1.a) and (2.1.b), and obtain the Poisson-Boltzmann (PB) equation in the following form as:

Electrolyte layer

$$\frac{\partial^2 \psi}{\partial y^2} + \frac{\partial^2 \psi}{\partial z^2} = \kappa^2 \psi \tag{2.3.a}$$

Polyelectrolyte layer

$$\frac{\partial^2 \psi}{\partial y^2} + \frac{\partial^2 \psi}{\partial z^2} = \kappa^2 \psi - \kappa_p^2 \tag{2.3.b}$$

Here, $\kappa = \left[(2z^2 e^2 n_\infty)/(\varepsilon_0 \varepsilon_r k_B T)\right]^{1/2}$ and $\kappa_p = \left[(zZe^2 N)/(\varepsilon_0 \varepsilon_r k_B T)\right]^{1/2}$ are the Debye-Hückel (DH) parameters of electrolyte and polyelectrolyte layers, respectively (Chanda *et al.* 2014; Gaikwad *et al.* 2018). The inverse of these parameters signifies the Debye length of electrolyte and equivalent Debye length of the polyelectrolyte layer, respectively. It should be noted that the equations (2.3.a) and (2.3.b) are obtained by appealing to the DH approximation in this analysis, i.e., $\sinh[(\mp ze\psi)/(k_B T)] \approx (\mp ze\psi)/(k_B T)$, as the potential at the walls is considered to be less than 25mV (Masliyah & Bhattacharjee 2006; R. J. Hunter 1981). Below, we non-dimensionalize equations (2.3.a) and (2.3.b) using the reference parameters: $l_{ref} = H$ and $\psi_{ref} = (k_B T)/(ze)$. The dimensionless forms of these equations read as:

Electrolyte layer

$$\frac{\partial^2 \bar{\psi}}{\partial \bar{y}^2} + \frac{\partial^2 \bar{\psi}}{\partial \bar{z}^2} = \bar{\kappa}^2 \bar{\psi} \tag{2.4.a}$$

Polyelectrolyte layer

$$\frac{\partial^2 \bar{\psi}}{\partial y^2} + \frac{\partial^2 \bar{\psi}}{\partial z^2} = \bar{\kappa}^2 \bar{\psi} - \bar{\kappa}_p^2 \qquad (2.4.b)$$

### 2.3   Rotational hydrodynamics coupled with PEL modulated electrostatics

In the present study, the primary flow $u(y,z)$ is driven by the combined influences of the pressure gradient and the electroosmotic force. In contrast, the secondary flow, depicted by $v(y,z)$ and $w(y,z)$ velocity components, is developed by the Coriolis force. The pressure gradient in x-direction corresponds to the gyrostatic pressure acting on the fluid, whereas the mass conservation constraint together with the Coriolis force $(\mathbf{\Omega} \times \mathbf{u})$ gives rise to the pressure gradients in the $y$ and $z$ directions (Ng & Qi 2015). If the effect of gravity is ignored, the governing equations for both the electrolyte layer and polyelectrolyte layer in the rotating frame of reference can be written as (Ng & Qi 2015):

Momentum conservation equation

Electrolyte layer

$$\rho\left(\frac{\partial \mathbf{u}}{\partial t} + (\mathbf{u}.\nabla)\mathbf{u} + 2(\mathbf{\Omega} \times \mathbf{u})\right) = -\nabla \mathbf{p} + \nabla.(\eta \mathbf{D}) + \rho_e \mathbf{E} \qquad (2.5.a)$$

Polyelectrolyte layer

$$\rho\left(\frac{\partial \mathbf{u}}{\partial t} + (\mathbf{u}.\nabla)\mathbf{u} + 2(\mathbf{\Omega} \times \mathbf{u})\right) = -\nabla \mathbf{p} + \nabla.(\eta \mathbf{D}) + \rho_e \mathbf{E} - \mu_c \mathbf{u} \qquad (2.5.b)$$

Continuity equation

$$\nabla.\mathbf{u} = 0 \qquad (2.6)$$

Note that in equations (2.5) and (2.6), $\mathbf{u} = (u, v, w)$ is the velocity vector in the rotating frame, $\mathbf{D} = (\nabla \mathbf{u} + \nabla \mathbf{u}^T)/2$ is the deformation rate tensor, $\eta$ is the viscosity, $\rho$ is the density of electrolyte, $\mathbf{p}$ is the gyrostatic pressure field, and $\mu_c$ signifies Darcy's frictional drag parameter. The term $\mu_c \mathbf{u}$ appearing in equation (2.5.b) represents the frictional drag force offered by the polyelectrolyte layer to the transport of electrolyte. It may be mentioned here that this frictional resistance depends on the grafting density and the thickness of the polyelectrolyte layer (Chanda *et al.* 2014; Chen & Das 2016; Das *et al.* 2015).

We further non-dimensionalize governing equations (2.5) and (2.6) using following reference and dimensionless parameters: $u_{ref} = u_{HS} = (-\varepsilon \zeta E)/\eta$, $l_{ref} = H$, $p_{ref} = (\eta u_{HS})/H$, $t_{ref} = H/u_{HS}$, $\text{Re} = (\rho u_{HS} H)/\eta$, $\text{Re}_\omega = (2\Omega \rho H^2)/\eta$ and $\alpha^2 = (\mu_c H^2)/\eta$ (Kaushik *et al.* 2017a, 2017b; Kaushik & Chakraborty 2017; Ng & Qi 2015). Using these parameters, the dimensionless transport equations can be written as:

Momentum conservation equation

Electrolyte layer

$$\text{Re}\left[\frac{\partial \bar{\mathbf{u}}}{\partial \bar{t}} + (\bar{\mathbf{u}}.\bar{\nabla})\bar{\mathbf{u}}\right] + (\text{Re}_\omega \times \bar{\mathbf{u}}) = -\bar{\nabla}\bar{\mathbf{p}} + \bar{\nabla}^2\bar{\mathbf{u}} + \bar{\kappa}^2\bar{\psi} \qquad (2.7.\text{a})$$

Polyelectrolyte layer

$$\text{Re}\left[\frac{\partial \bar{\mathbf{u}}}{\partial \bar{t}} + (\bar{\mathbf{u}}.\bar{\nabla})\bar{\mathbf{u}}\right] + (\text{Re}_\omega \times \bar{\mathbf{u}}) = -\bar{\nabla}\bar{\mathbf{p}} + \bar{\nabla}^2\bar{\mathbf{u}} + \bar{\kappa}^2\bar{\psi} - \alpha^2\bar{\mathbf{u}} \qquad (2.7.\text{b})$$

Continuity equation

$$\bar{\nabla}.\bar{\mathbf{u}} = 0 \qquad (2.8)$$

It is worth to mention here that the inverse of rotational Reynolds number $\text{Re}_\omega$ signifies the definition of Ekman number, which is nothing but the ratio of viscous force to the Coriolis force. Particularly for this study, the range of Ekman number is moderate as both the Coriolis force and viscous force are significant in this analysis. Also, in the present study, considering low $\text{Re}(\ll 1)$, we neglect the influence of inertia on the underlying analysis. Owing to these approximations, and for the steady flow analysis, equations (2.7.a) and (2.7.b) can be rewritten as given below (Ng & Qi 2015):

Electrolyte layer

$$\text{Re}_\omega \times \bar{\mathbf{u}} = -\bar{\nabla}\bar{\mathbf{p}} + \bar{\nabla}^2\bar{\mathbf{u}} + \bar{\kappa}^2\bar{\psi} \qquad (2.9.\text{a})$$

Polyelectrolyte layer

$$\text{Re}_\omega \times \bar{\mathbf{u}} = -\bar{\nabla}\bar{\mathbf{p}} + \bar{\nabla}^2\bar{\mathbf{u}} + \bar{\kappa}^2\bar{\psi} - \alpha^2\bar{\mathbf{u}} \qquad (2.9.\text{b})$$

Finally, the system of governing equations for the present study can be written as:

Momentum conservation equation

Electrolyte layer

*x momentum*

$$-\text{Re}_\omega \bar{v} = -\frac{\partial \bar{p}}{\partial \bar{x}} + \frac{\partial^2 \bar{u}}{\partial \bar{y}^2} + \frac{\partial^2 \bar{u}}{\partial \bar{z}^2} + \bar{\kappa}^2\bar{\psi} \qquad (2.10.\text{a})$$

*y momentum*

$$\text{Re}_\omega \bar{u} = -\frac{\partial \bar{p}}{\partial \bar{y}} + \frac{\partial^2 \bar{v}}{\partial \bar{y}^2} + \frac{\partial^2 \bar{v}}{\partial \bar{z}^2} \qquad (2.10.\text{b})$$

*z momentum*

$$0 = -\frac{\partial \bar{p}}{\partial \bar{z}} + \frac{\partial^2 \bar{w}}{\partial \bar{y}^2} + \frac{\partial^2 \bar{w}}{\partial \bar{z}^2} \qquad (2.10.\text{c})$$

Polyelectrolyte layer

*x momentum*

$$-\text{Re}_\omega \bar{v} = -\frac{\partial \bar{p}}{\partial \bar{x}} + \frac{\partial^2 \bar{u}}{\partial \bar{y}^2} + \frac{\partial^2 \bar{u}}{\partial \bar{z}^2} + \bar{\kappa}^2 \bar{\psi} - \alpha^2 \bar{u} \qquad (2.11.\text{a})$$

*y momentum*

$$\text{Re}_\omega \bar{u} = -\frac{\partial \bar{p}}{\partial \bar{y}} + \frac{\partial^2 \bar{v}}{\partial \bar{y}^2} + \frac{\partial^2 \bar{v}}{\partial \bar{z}^2} - \alpha^2 \bar{v} \qquad (2.11.\text{b})$$

*z momentum*

$$0 = -\frac{\partial \bar{p}}{\partial \bar{z}} + \frac{\partial^2 \bar{w}}{\partial \bar{y}^2} + \frac{\partial^2 \bar{w}}{\partial \bar{z}^2} - \alpha^2 \bar{w} \qquad (2.11.\text{c})$$

Continuity equation

$$\frac{\partial \bar{v}}{\partial \bar{y}} + \frac{\partial \bar{w}}{\partial \bar{z}} = 0 \qquad (2.12)$$

For brevity in the presentation, henceforth, we drop the overbars from the dimensionless variables and use a symbol $G$ for $-\partial \bar{p}/\partial \bar{x}$.

## 3 Variational calculus

### 3.1 *Euler-Lagrange form of governing equations and boundary conditions*

We use the Ritz technique, which is consistent with the variational calculus method, to solve the transport equations of the present flow configuration [(2.4), (2.10)-(2.12)]. Following this method, we first compose the total functional using the Euler-Lagrange forms of these equations. For this task, we multiply these equations with their corresponding test functions, integrate over the control volume, and then add them together to get the total functional of the system (Reddy & Rasmussen 1982; Rektorys 1977). The minimization of this functional with respect to the Ritz coefficients (introduced in subsection 3.3) gives the approximate solution to the problem. The derivation of the functional over the total domain (cross-section of the channel), instead of solving differential equations [(2.4), (2.10)- (2.12)] for every sub-domain if any in the flow configuration, sets uniformity in the solution over the entire cross-section of the channel (Sadeghi 2018; Sadeghi *et al.* 2020). This advantageous feature prevails even with the consideration of the polyelectrolyte layer, which forms a separate sub-domain, in the underlying analysis [(2.10)-(2.11)] (Sadeghi 2018; Sadeghi *et al.* 2019, 2020). In such a case, a single functional is derived for the entire cross-section, and the Darcy drag $\left(\alpha^2 \mathbf{u}\right)$ and the term $\kappa_p^2$ is made zero for the electrolyte layer [see equation 2.10(a)-(c)]. The final set of governing equations, which are needed to be solved over the complete channel cross-section, can be written as:

Poisson-Boltzmann equation

$$\frac{\partial^2 \psi}{\partial y^2} + \frac{\partial^2 \psi}{\partial z^2} = \kappa^2 \psi - \kappa_p^2 \tag{3.1}$$

Momentum conservation equation
*x momentum*

$$-\text{Re}_\omega v = \frac{\partial^2 u}{\partial y^2} + \frac{\partial^2 u}{\partial z^2} + G + \kappa^2 \psi - \alpha^2 u \tag{3.2.a}$$

*y momentum*

$$\text{Re}_\omega u = -\frac{\partial p}{\partial y} + \frac{\partial^2 v}{\partial y^2} + \frac{\partial^2 v}{\partial z^2} - \alpha^2 v \tag{3.2.b}$$

*z momentum*

$$0 = -\frac{\partial p}{\partial z} + \frac{\partial^2 w}{\partial y^2} + \frac{\partial^2 w}{\partial z^2} - \alpha^2 w \tag{3.2.c}$$

Continuity equation

$$\frac{\partial v}{\partial y} + \frac{\partial w}{\partial z} = 0 \tag{3.3}$$

Below, we write the boundary conditions that accompany the set of governing equations mentioned above $(3.1)-(3.3)$ as:

For the potential distribution

$$\left.\frac{\partial \psi}{\partial (y,z)}\right|_{y,z=\pm 1} = 0 \tag{3.4.a}$$

$$\left.\frac{\partial \psi}{\partial (y,z)}\right|_{y,z=0} = 0 \tag{3.4.b}$$

The first and second boundary conditions in (3.4) depict the no-flux boundary condition at the walls and the center (i.e., at y and z-axis), respectively.

For the velocity distribution

$$\left.(u,v,w)\right|_{y,z=\pm 1} = 0 \tag{3.5.a}$$

$$\left.\frac{\partial (u,v)}{\partial (y,z)}\right|_{y,z=0} = 0 \tag{3.5.b}$$

$$\left. w \right|_{y,z=0} = 0 \tag{3.5.c}$$

The first boundary condition (3.5.a) depicts the no-slip boundary condition at the walls of the channel for all velocity components $u$, $v$ and $w$. The second boundary condition (3.5.b) is considered for the symmetric nature of the profiles of both $u$ and $v$ velocity components at $y$ and $z$-axis. The last boundary condition (3.5.c) stands for the anti-symmetric nature of $w$ velocity at $y$ and $z$-axis.

## 3.2 Formulation of functional

Let $\delta\psi$, $\delta u$, $\delta v$, $\delta w$ and $\delta p$ be the variations in $\psi$, $u$, $v$, $w$ and $p$ variables, where $\delta\psi, \delta u, \delta v, \delta w, \delta p \in H_0^2(\Omega_b)$ and $\Omega_b$ is the domain with boundary $\partial\Omega_b$. In the process of selection of these variations, it is necessary to take into account the inter-dependency of these variations ($\delta\psi$, $\delta u$, $\delta v$, $\delta w$ and $\delta p$) on each other as the present problem deals with the two-way coupled system of equations [(3.1)-(3.3)]. In the rotational flow dynamics, an increment in $v$, $w$ and $p$ (secondary flow) leads to a reduction in the variable $u$ (axial flow) due to Coriolis force and the induced pressure gradients in y and z directions. Whereas the potential distribution remains unaffected as it does not depend on the flow velocities due to insignificant ionic Peclet number (Bandopadhyay *et al.* 2016; Gaikwad *et al.* 2020; Zaccone *et al.* 2009). Therefore, accounting all these aspects, we impose following sign convections on the aforementioned variations as,

$$\delta\psi = +\text{ve},\ \delta u = +\text{ve},\ \delta v = -\text{ve},\ \delta w = -\text{ve},\ \delta p = -\text{ve}$$

To formulate the functional, we use the definition of the first variation for any function $F(y, z, \phi, \phi_y, \phi_z) = 0$, where $\phi_y = \partial\phi/\partial y$ and $\phi_z = \partial\phi/\partial z$,

$$\delta I(\phi) = \int_{\Omega_b}\left[\frac{\partial F}{\partial \phi} - \frac{\partial}{\partial x}\left(\frac{\partial F}{\partial \phi_y}\right) - \frac{\partial}{\partial y}\left(\frac{\partial F}{\partial \phi_z}\right)\right]\delta\phi\, dydz \tag{3.6}$$

where $\delta I(\phi)$ is the first variation of functional $I(\phi)$ and $\delta\phi$ is the variation of $\phi(y, z)$. Applying the above definition (3.6) to the equations (3.1) to (3.3), we get,

$$\delta I(\psi) = \int_{\Omega_b}\left[\frac{\partial^2\psi}{\partial y^2} + \frac{\partial^2\psi}{\partial z^2} - \kappa^2\psi + \kappa_p^2\right]\delta\psi\, dydz \tag{3.7.a}$$

$$\delta I(u) = \int_{\Omega_b}\left[\frac{\partial^2 u}{\partial y^2} + \frac{\partial^2 u}{\partial z^2} + \kappa^2\psi - \alpha^2 u + G + \text{Re}_\omega v\right]\delta u\, dydz \tag{3.7.b}$$

$$\delta I(v) = \int_{\Omega_b}\left[\frac{\partial^2 v}{\partial y^2} + \frac{\partial^2 v}{\partial z^2} - \frac{\partial p}{\partial y} - \alpha^2 v - \text{Re}_\omega u\right](-\delta v)\, dydz \tag{3.7.c}$$

$$\delta I(w) = \int_{\Omega_b}\left[\frac{\partial^2 w}{\partial y^2} + \frac{\partial^2 w}{\partial z^2} - \frac{\partial p}{\partial z} - \alpha^2 w\right](-\delta w)\, dydz \tag{3.7.d}$$

$$\delta I(p) = \int_{\Omega_b} \left[ \frac{\partial v}{\partial y} + \frac{\partial w}{\partial z} \right](-\delta p)\,dydz \qquad (3.7.e)$$

In the above equations (3.7), the integrand of each integral represents the energy or work associated with the corresponding variable. The integrand of first integral [Eq. (3.7.a)] represents the electrical energy in the domain $\Omega_b$, whereas the integrands of equations (3.7.b)-(3.7.d) represent the kinetic energy associated with the velocities $u$, $v$ and $w$ (Reddy 1986; Reddy & Rasmussen 1982). In equation (3.7.e), the integrand represents the flow work done by the pressure to conserve the mass of the system (Reddy 1986; Reddy & Rasmussen 1982). Now, we perform the integrations in equations (3.7) and add them together to get the functional $\Pi$ as,

$$\Pi = \int_{\Omega_b} \left[ \left(\frac{\partial \psi}{\partial y}\right)^2 + \left(\frac{\partial \psi}{\partial z}\right)^2 + \left(\frac{\partial u}{\partial y}\right)^2 + \left(\frac{\partial u}{\partial z}\right)^2 - \left(\frac{\partial v}{\partial y}\right)^2 - \left(\frac{\partial v}{\partial z}\right)^2 - \left(\frac{\partial w}{\partial y}\right)^2 - \left(\frac{\partial w}{\partial z}\right)^2 \right. \\ \left. -\left(2\kappa^2 \psi - 2G\right)u - 2\mathrm{Re}_\omega uv + \alpha^2\left(u^2 - v^2 - w^2\right) + 2p\left(\frac{\partial v}{\partial y} + \frac{\partial w}{\partial z}\right) + \kappa^2 \psi^2 - 2\kappa_p^2 \psi \right] dydz \qquad (3.8)$$

### 3.3    *Extremization of functional: Ritz method*

Now, we obtain the solution to the problem using the Ritz method in which the variables $u$, $v$, $w$, $p$ and $\psi$ are approximated by the corresponding basis functions. Note that the selection of suitable basis functions is integral as well as an essential part of this method. We stick to the following basics for selecting these functions in the present endeavour, i.e., the basis functions should satisfy the boundary conditions (3.4)-(3.5) as well as the symmetric or asymmetric nature of the variables (Ng & Qi 2015). In this study, the $u$ and $v$ velocity components, as well as the potential $\psi$ has symmetric nature about y and z-axis. In contrast, the velocity $w$ has an anti-symmetric nature about y and z-axis. Also, the basis function chosen for pressure should maintain orthogonality with functions that have already been selected for $v$ and $w$ velocity components. Accounting all these aspects, we consider the following test functions for the functional $\Pi$ as:

$$\psi(y,z) = \sum_{i=1}^{M} \psi_{bi} \psi_{fi} = \sum_{i=1}^{M} \psi_{bi} \cos(l_{\psi i} \pi y) \cos(m_{\psi i} \pi z) \qquad (3.9.a)$$

$$u(y,z) = \sum_{i=1}^{M} u_{bi} u_{fi} = \sum_{i=1}^{M} u_{bi} \cos\left[(2l_{ui}+1)\pi \frac{y}{2}\right] \cos\left[(2m_{ui}+1)\pi \frac{z}{2}\right] \qquad (3.9.b)$$

$$v(y,z) = \sum_{i=1}^{M} v_{bi} v_{fi} = \sum_{i=1}^{M} v_{bi} \cos\left[(2l_{vi}+1)\pi \frac{y}{2}\right] \cos\left[(2m_{vi}+1)\pi \frac{z}{2}\right] \qquad (3.9.c)$$

$$w(y,z) = \sum_{i=1}^{M} w_{bi} w_{fi} = \sum_{i=1}^{M} w_{bi} \sin\left[(l_{wi}+1)\pi y\right] \sin\left[(m_{wi}+1)\pi z\right] \qquad (3.9.\text{d})$$

$$p(y,z) = \sum_{i=1}^{M} p_{bi} p_{fi} = \sum_{i=1}^{M} p_{bi} \left( \begin{array}{l} \left( \sin\left[(l_{pi}+1)\pi y\right] \cos\left[(2m_{pi}+1)\pi \dfrac{z}{2}\right] \right) \\ + p_c \left( \sin\left[(2l_{pi}+1)\pi \dfrac{y}{2}\right] \cos\left[(2m_{pi}+1)\pi \dfrac{z}{4}\right] \right) \end{array} \right) \qquad (3.9.\text{e})$$

In equations (3.9), the coefficients $\psi_{bi}$, $u_{bi}$, $v_{bi}$, $w_{bi}$ and $p_{bi}$ are the Ritz coefficients with respect to which functional (3.8) needs to be minimized. For minimization, we set,

$$\frac{\partial \Pi}{\partial \psi_{bi}} = 0 \text{ where } \forall \psi_{bi} \in [1, M] \qquad (3.10.\text{a})$$

$$\frac{\partial \Pi}{\partial u_{bi}} = 0 \text{ where } \forall u_{bi} \in [1, M] \qquad (3.10.\text{b})$$

$$\frac{\partial \Pi}{\partial v_{bi}} = 0 \text{ where } \forall v_{bi} \in [1, M] \qquad (3.10.\text{c})$$

$$\frac{\partial \Pi}{\partial w_{bi}} = 0 \text{ where } \forall w_{bi} \in [1, M] \qquad (3.10.\text{d})$$

$$\frac{\partial \Pi}{\partial p_{bi}} = 0 \text{ where } \forall p_{bi} \in [1, M] \qquad (3.10.\text{e})$$

where $l_{\psi i}, l_{ui}, l_{vi}, l_{wi}, l_{pi} \in [0, m]$ and $m_{\psi i}, m_{ui}, m_{vi}, m_{wi}, m_{pi} \in [0, m]$; hence, $M = (m+1) \times (m+1)$. After minimizing the functional (3.8), we get,

$$\sum_{j=1}^{M} \psi_{bj} \int_{\Omega_b} \left[ \frac{\partial \psi_{fj}}{\partial y} \frac{\partial \psi_{fi}}{\partial y} + \frac{\partial \psi_{fj}}{\partial z} \frac{\partial \psi_{fi}}{\partial z} + \kappa^2 \psi_{fj} \psi_{fi} \right] dydz = \int_{\Omega_b} \kappa_p^2 \psi_{fi} dydz \qquad (3.11.\text{a})$$

$$\sum_{j=1}^{M} u_{bj} \int_{\Omega_b} \left[ \frac{\partial u_{fj}}{\partial y} \frac{\partial u_{fi}}{\partial y} + \frac{\partial u_{fj}}{\partial z} \frac{\partial u_{fi}}{\partial z} + \alpha^2 u_{fj} u_{fi} \right] dydz = \int_{\Omega_b} \left[ \kappa^2 \left( \sum_{k=1}^{M} \psi_{bk} \psi_{fk} \right) + G \right] u_{fi} dydz$$
$$+ \int_{\Omega_b} \left[ \text{Re}_\omega \left( \sum_{k=1}^{M} v_{bk} v_{fk} \right) \right] u_{fi} dydz \qquad (3.11.\text{b})$$

$$\sum_{j=1}^{M} v_{bj} \int_{\Omega_b} \left[ \frac{\partial v_{fj}}{\partial y} \frac{\partial v_{fi}}{\partial y} + \frac{\partial v_{fj}}{\partial z} \frac{\partial v_{fi}}{\partial z} + \alpha^2 v_{fj} v_{fi} \right] dydz = \int_{\Omega_b} \left[ -\left( \sum_{k=1}^{M} p_{bk} \frac{\partial p_{fk}}{\partial y} \right) \right] v_{fi} dydz$$
$$- \int_{\Omega_b} \left[ \text{Re}_\omega \left( \sum_{k=1}^{M} u_{bk} u_{fk} \right) \right] v_{fi} dydz \qquad (3.11.\text{c})$$

$$\sum_{j=1}^{M} w_{bj} \int_{\Omega_b} \left[ \frac{\partial w_{fj}}{\partial y} \frac{\partial w_{fi}}{\partial y} + \frac{\partial w_{fj}}{\partial z} \frac{\partial w_{fi}}{\partial z} + \alpha^2 w_{fj} w_{fi} \right] dydz = \int_{\Omega_b} -\left( \sum_{k=1}^{M} p_{bk} \frac{\partial p_{fk}}{\partial z} \right) w_{fi} dydz \qquad (3.11.\text{d})$$

$$\int_{\Omega_b} p_{fi}\left[\left(\sum_{j=1}^{M} v_{bj}\frac{\partial v_{fj}}{\partial y}\right)+\left(\sum_{j=1}^{M} w_{bj}\frac{\partial w_{fj}}{\partial z}\right)\right]dydz = 0 \qquad (3.11.e)$$

In equation (3.9.e), the basis function of pressure $p_{fi}$ has an additional term $p_2$ $\left(=\sin\left[(2l_{pi}+1)\pi y/2\right]\cos\left[(2m_{pi}+1)\pi z/4\right]\right)$ with a coefficient $p_c$, which assists in maintaining the mass conservation in the domain. Otherwise, the boundary conditions imposed by function $p_1$ $\left(=\sin\left[(l_{pi}+1)\pi y\right]\cos\left[(2m_{pi}+1)\pi z/2\right]\right)$ lead to divergence in the continuity.

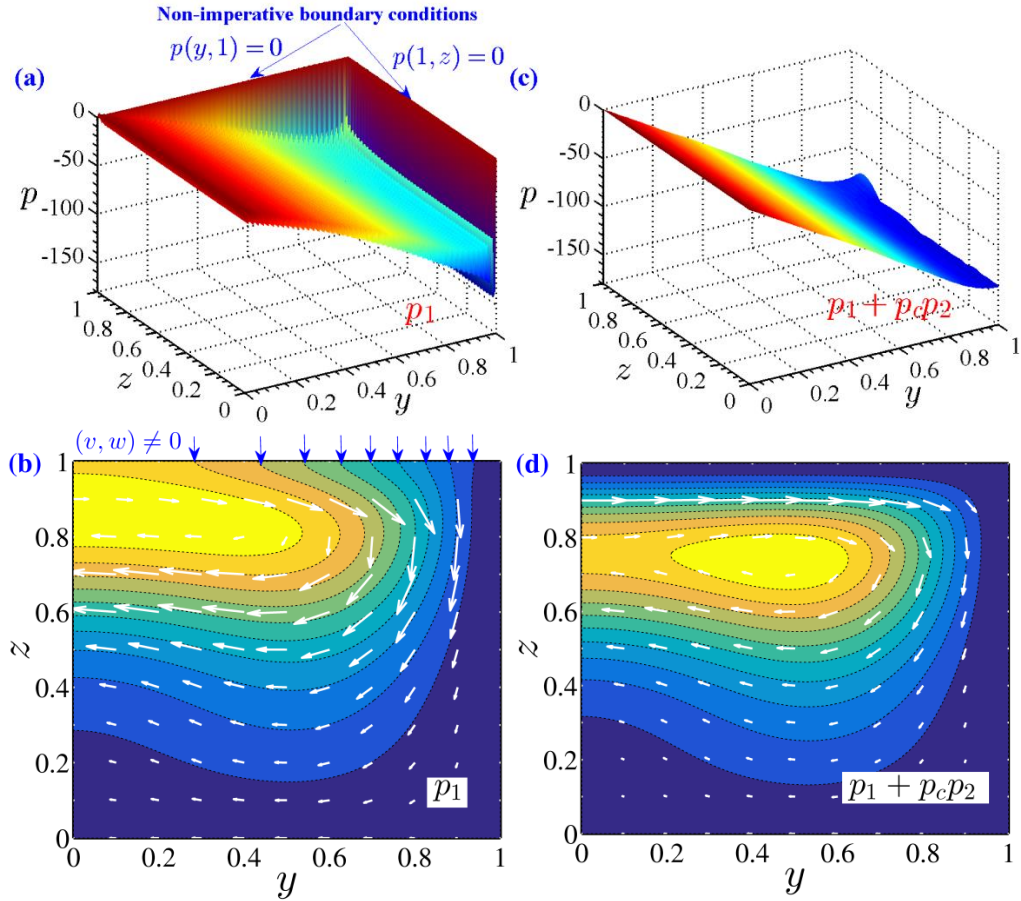

FIGURE 2 (color online): Surface plots of pressure distribution ((a) and (c)) and contour plots of secondary flow ((b) and (d)) for basis functions $p_1$ and $p_1 + p_c p_2$. The first column ((a) and (b)) depicts the case of basis function $p_1$ and second column ((c) and (d)) depicts the case of $p_1 + p_c p_2$. The contour plots of stream function ((b) and (d)) give the inference of the divergence in the mass conservation in the channel. The parameters considered to obtain these plots are: $\kappa_p = 10$, $d = 0.3$, $\alpha = 5$, $G = 0.5$, $\text{Re}_\omega = 100$, $\kappa = 12$ and $l_i \times m_i = 100 \times 100$.

The effect of these functions *viz.* $p_1$ and $p_1 + p_c p_2$ on the pressure distribution and the corresponding secondary flow development is shown in figure 2. The contour plots of the secondary flow in figures 2(b) and 2(d) help to analyze the divergence in the mass conservation $(\nabla \cdot \mathbf{u} = 0)$. In case of divergence, the walls of the channel do not imitate the streamlines. Thus, the streamlines of the secondary flow pass through the walls, signifying the transaction of fluid mass across the walls of the channel. In the present study, this phenomenon takes place for the basis function $p_1$ as evident in figure 2(b). The non-imperative boundary conditions at the walls $p(y,1) = p(1,z) = 0$ imposed by the basis function $p_1$ (see figure 2(a)) lead to divergence in $\nabla \cdot \mathbf{u} = 0$ as shown in figure 2(b). In contrast, the additional term $p_c p_2$ in the basis function $p_1 + p_c p_2$ leads to the conservation of mass in the channel, as depicted by the parallel or non-crossing streamlines at the walls of the channel in figure 2(d). The modified pressure distribution, according to the basis function, $p_1 + p_c p_2$ is shown in figure 2(c). Note that the value of $p_c$ is based on the order of the continuity equation $O[\nabla \cdot \mathbf{u}]$. For a given set of other physical parameters, the value of $p_c$ is obtained as ~2.4 using the trial and error method.

The distribution of electrostatic potential $\psi$, velocity field $\mathbf{u}$, and pressure $p$ in $y-z$ plane is obtained by solving equations (3.11) with the test functions (3.9). For a concise presentation, we do not outline the intermediate steps here. Instead, the detailed solution procedure to solve equations (3.11) is given in the supporting information part of the paper (section S2).

## 4 The physical significance of model parameters

In rotational microflows, the thickness of the Ekman layer varies between $O[10^1-10^2]\,\mu\text{m}$ for the rotational speed of centrifuge up to 1000 rpm (Optima L-90K, Beckman Coulter, Inc.). The length scale of microchannels typically varies within the range $O[10^0-10^2]\,\mu\text{m}$ (Duffy et al. 1999). Accordingly, the range of dimensionless rotational speed $\text{Re}_\omega\left[=(2\Omega\rho H^2)/\eta\right]$ can be estimated as $O[10^0-10^2]$ (Duffy et al. 1999). Here, the universal solvent 'water' is considered as the working fluid, i.e., $\rho = 1000\,\text{kg/m}^3$ and $\mu = 10^{-3}\,\text{Pa.s}$. Besides these rotation-dependent parameters, the other parameters which significantly modulate the flow dynamics in the present configuration are the DH parameter of EL $(\kappa)$ and PEL $(\kappa_p)$, PEL thickness $(d)$, and drag parameter $(\alpha)$. We would like to mention here that the range of PEL thickness $d$ (= 0.05-0.3) and drag parameter $\alpha$ (=1-20) considered in the present study are consistent with the reported values in the literature (Gaikwad *et al.* 2018; Kaushik *et al.* 2017a). To estimate the range of the DH parameter ($\kappa$ or $\kappa_p$), we resort to the electrolyte concentration and the thermodynamic state properties of the system. Under the standard laboratory conditions, the thickness of the electrical double layer varies in the range ~ 10-300 nm, if the concentration of the symmetric electrolyte (such as KCl, NaCl) is varied between 1-0.001mM (Masliyah & Bhattacharjee 2006). Considering this range, for the typical length scale of the microchannel (10-$10^2 \mu$m as mentioned before), the value of the dimensionless DH parameter may fall in the range 1 to 100. It is important to mention here that the aforementioned values of the basic and derived parameters are typical to the microfluidic setup and commonly used in the literature as well (Das *et al.* 2015; Gaikwad *et al.* 2018; Kaushik *et al.* 2017a).

## 5  Model benchmarking

Before going to discuss the results at hand, we make an effort to benchmark our theoretical model in this section. In doing so, we take advantage of full-scale numerical simulations as elaborated next. Besides, we also discuss the consistency of the velocity profiles of the present problem from different perspectives, as reported in the literature.

### 5.1  *Comparison with Finite Element Method results*

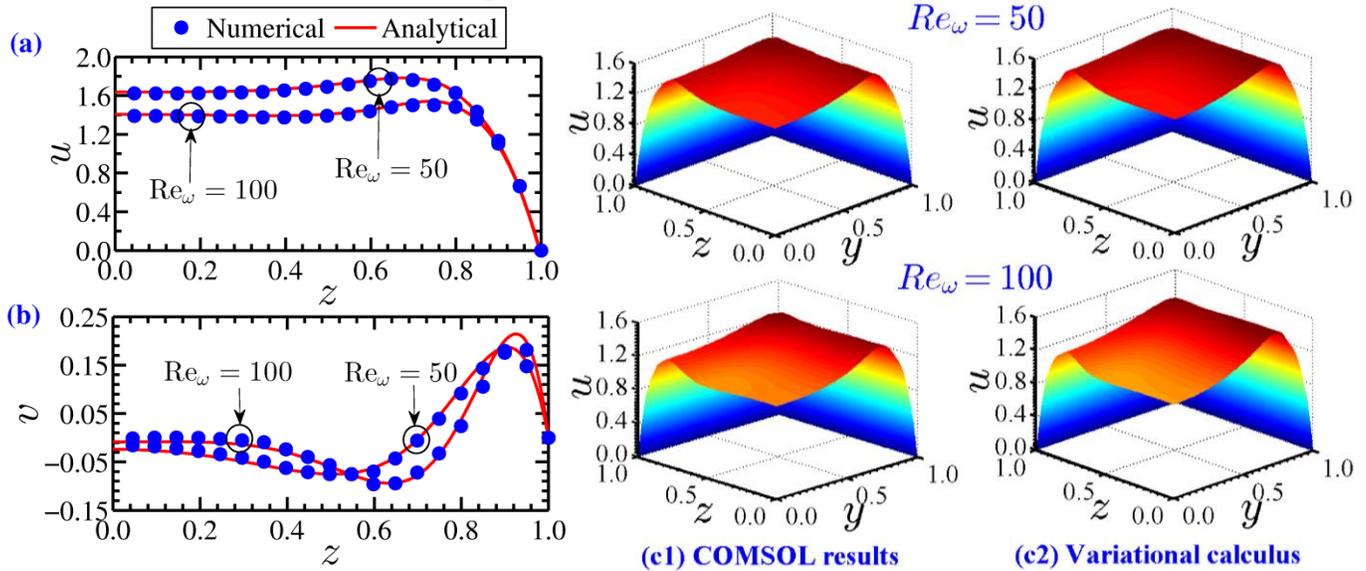

FIGURE 3 (color online). Validation plots of the variational calculus method for different rotational speeds $Re_\omega = 50$ and $Re_\omega = 100$. **Validation through line plots:** Comparison through the line plots (in (a) and (b)) of $u$ and $v$ velocity profile obtained at $y = 0$. **Validation through surface plots:** Comparison through the surface plots (in (c)) of $u$-velocity profile. The other parameters considered for this analysis are: $\kappa_p = 10$, $d = 0.2$, $\alpha = 5$, $G = 0.5$ and $\kappa = 12$.

In figure 3(a)-(c), we compare the results obtained from the variational calculus method with those obtained from the finite element method (FEM) based COMSOL Multiphysics™ solver under identical condition, as mentioned in the caption. The plots in figures 3(a) and 3(b) show 1D profiles of $u$-velocity and $v$-velocity obtained at $y = 0$, respectively, whereas the surface plots in figure 3(c) map the 3D velocity profiles, obtained for rotational speeds $\text{Re}_\omega = 50$ and $\text{Re}_\omega = 100$. Note that in variational calculus method, the number of basis functions selected to compute the velocity distribution corresponds to the number of indices. For the present analysis, we consider the following set as $l_i = 100$ and $m_i = 100$ for each variable $u$, $v$, $w$, $p$ and $\psi$.

It may be mentioned here that for the validation plots depicted above in figure 3(a)-(c), we use the creeping flow model for hydrodynamics and the Poisson equation for potential distribution in COMSOL solver. To obtain the fully developed flow, we use a considerably long channel of 3D rectangular geometry with length to height (or width) ratio $O\left[10^2\right]$ in the COMSOL framework. It can be observed in figure 3(a)-(c) that the velocity profiles obtained from our theoretical model accurately mimic the results of the COMSOL solver. Quite notably, it is evident from figures 3(a) and 3(c) that the peak value of the axial velocity occurs at a distance approximately equal to $2\text{Re}_\omega^{-1/2}$ (Kheshgi & Scriven 1985; Speziale 1982) from the horizontal wall of the channel. This observation is consistent with the theory of the linear Ekman layer, elaborated in Hart 1971 (Hart 1971).

### 5.2 *Consistency of the flow dynamics from different perspectives*

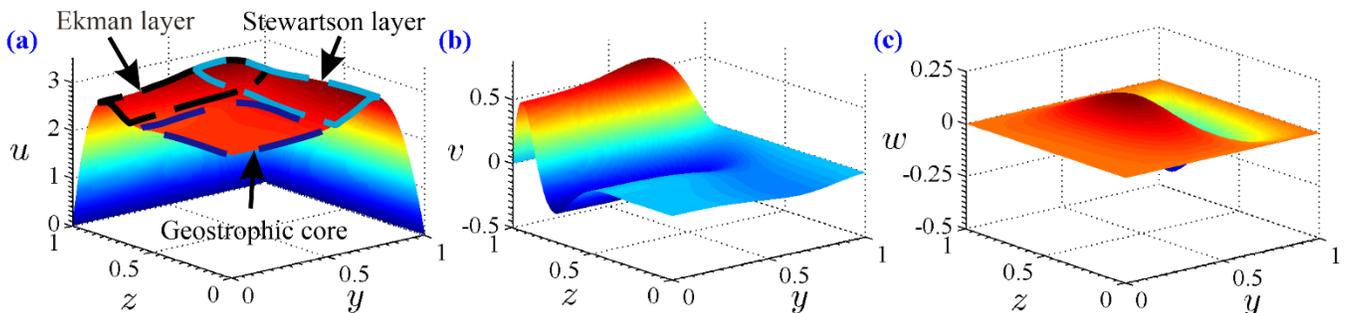

FIGURE 4 (color online). Velocity profiles of $u$, $v$ and $w$ velocity, obtained for $\kappa_p = 10$, $d = 0.3$, $\alpha = 5$, $G = 0.5$, $\text{Re}_\omega = 150$ and $\kappa = 12$.

In the present set up, the flow is driven by the combined influences of the gyrostatic pressure gradient and the electroosmotic force. In the asymptotic limit of geostrophic plug flow $(\text{Ro} \sim 0)$, as depicted in figure 4(a), a plug-like axial velocity profile appears with a geostrophic core at the center, Ekman layer at the upper wall (i.e., at $z = +1$) and Stewartson layer at the left wall of the channel (i.e., at $y = +1$). As seen in figure 4(a), the magnitude of axial velocity $(u)$ increases from zero to its peak within both the Ekman layer and the Stewartson layer, while it assumes a nearly constant trend in the geostrophic core. To be precise, this typical flow behavior is in complete agreement with the well-established Taylor-Proudman theorem, i.e., $\partial_z(u,v) = 0$ and $(u,v).(\partial_x p, \partial_y p) = 0$ (Pedlosky 1987). Analogous to this theorem and upon satisfying the constraint of mass conservation principle $(\nabla \cdot \mathbf{u} = 0)$, we obtain the profiles of $v$ and $w$ velocities as depicted in figures 4(b) and 4(c), respectively. Note that the flat depression in the velocities $u$ and $v$ at the inviscid interior emanates owing to the definite balance between the Coriolis force and the pressure gradient for $\text{Re}_\omega \gg O(10^0)$ (Kheshgi & Scriven 1985; Speziale 1982).

Notable insights, as observed from figures 3 and 4, are in support of the effectiveness of the variational calculus method in predicting the flow dynamics in grafted PEL modulated rotational fluidics. Having established the accuracy of the present theoretical model, we now proceed for the systematic discussion of the underlying transport features, as depicted in the forthcoming sections.

## 6  Effect of PEL on rotational electrohydrodynamics

It may be mentioned here that, in rotational microfluidics, the Coriolis force adversely affects the deliverance of the prime objective of the micro total analysis systems $(\mu\text{TAS})$, i.e., augmented mixing without compromising the net throughput (Ng & Qi 2015). Although assisted by the electroosmotic effect, yet the rotational induced forces bring in a restriction on the underlying mixing performance as well as on the net throughput beyond certain rotational speeds. Therefore, ensuring these objectives with equal priority demands a severe challenge in the design of state of the art Lab-on-CD based microfluidic devices. To this end, the grafted PEL at the walls of the microfluidic channel has shown tremendous potential, attributed primarily to their inherent physicochemical characteristics to control the underlying transport. Considering this feature of PEL, we discuss, in this section, the flow dynamics in a rotating microfluidic channel, as influenced by the grafted PEL at the walls. Note that for the subsequent discussion, we consider a higher rotational speed, as manifested by a relatively larger $\text{Re}_\omega (\gg 1)$.

### 6.1 Primary flow

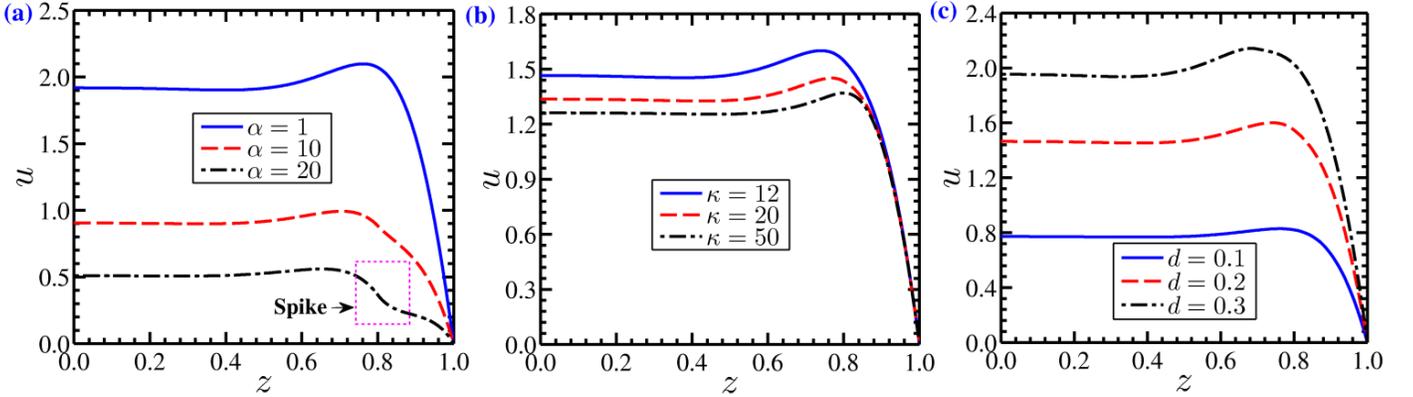

FIGURE 5(color online). Plots of axial velocity for the variation of drag parameter $\alpha$ (= 1, 10 and 20), DH parameter of EL $\kappa$ (= 12, 20 and 50) and PEL thickness $d$ (= 0.1, 0.2 and 0.3). The other parameters considered for this analysis are: (a) $\kappa_p = 10$, $d = 0.2$, $\kappa = 12$, $G = 0.5$, $\text{Re}_\omega = 100$; (b) $\kappa_p = 10$, $d = 0.2$, $\alpha = 5$, $G = 0.5$, $\text{Re}_\omega = 100$; and (c) $\kappa_p = 10$, $\alpha = 5$, $G = 0.5$, $\text{Re}_\omega = 100$ and $\kappa = 12$.

In figure 5(a)-(c), the plots show the influence of the physicochemical parameters of the PEL on the primary flow dynamics. For these plots, we consider three different values of (a) drag parameter $\alpha$ (= 1, 10 and 20), (b) DH parameter of EL $\kappa$ (= 12, 20 and 50), and (c) PEL thickness $d$ (= 0.1, 0.2 and 0.3). The plots in figures 6(a) and 6(b) show the variation in the potential distribution with a change in $\kappa$ and $d$ respectively. In figure 6(c), the plots depict the variation of electroosmotic force $F_e$ with $\kappa$. Note that the depicted variations in figure 6 are useful to correlate the distribution of axial flow velocity shown in figure 5. Also, the inferences of figure 6 will be referred to explain the results, as discussed in the forthcoming sections.

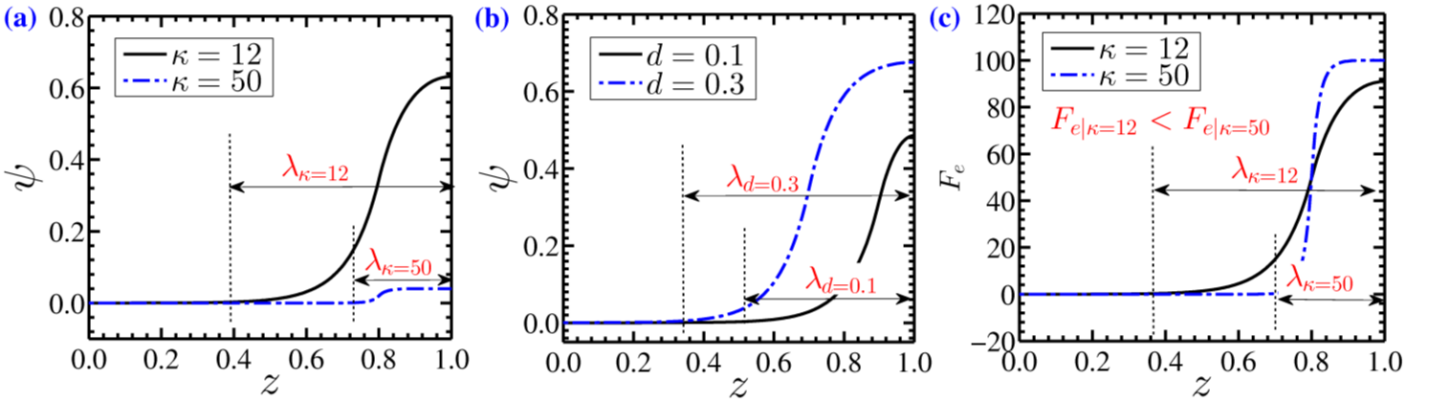

FIGURE 6 (color online): **Potential distribution:** (a)-(b) Plots showing the distribution of induced electrostatic potential $\psi$ for different values of DH parameter of EL $\kappa$ (= 12 and 50) and PEL thickness $d$ (= 0.1 and 0.3), respectively. **Variation of electroosmotic force with DH parameter of EL:** (c) Plots showing the variation of electroosmotic force ($F_e$) for different values of DH parameter of EL $\kappa$ (= 12 and 50). The other parameters considered are: (a) and (c) $d = 0.2$, $\kappa_p = 10$; (b) $\kappa = 12$, $\kappa_p = 10$.

We observe in figure 5 that the primary flow and the intensity of depression in the geostrophic core is comparatively higher for lower values of $\alpha$ (=1) and $\kappa$ (=12), and higher value of $d$ (=0.3); although the rotational speed is remaining unaltered. This variation in the intensity of depression in the geostrophic core is attributed to the effect of Coriolis force $(\text{Re}_\omega \times \mathbf{u})$, which not only varies with the rotational speed $\text{Re}_\omega$ but also depends on the magnitude of flow velocity $|\mathbf{u}|$ (Kheshgi & Scriven 1985; Ng & Qi 2015; Speziale 1982). In the present study, the magnitude of flow velocity largely depends on the electroosmotic effect and Darcy's frictional drag in PEL, thus is governed by the magnitude of pertinent parameters viz., $\alpha$, $\kappa$, and $d$. Note that $\alpha = 1$ signifies relatively lower drag force in the PEL region, whereas, for $\kappa = 12$ (figure 6(a)) and $d = 0.3$ (figure 6(b)), the electroosmotic effect becomes stronger. As shown in figure 6(a)-(c), for $\kappa = 12$ and $d = 0.3$, the stronger ionic interactions between PEL and electrolyte ions leads to an increasing magnitude of electrostatic potential $(\psi)$, which, in turn, enhances the electroosmotic effect. It is also observed in figure 6(a)-(c) that the thickness of EDL obtained for $\kappa = 12$ and $d = 0.3$ is higher than their respective counterparts, signifying a thicker influencing regime of the electroosmotic effect in the channel for the said values. A relatively higher magnitude of potential together with its influence up to a larger distance for $d = 0.3$, as seen in figure 6(b), is suggestive of achieving higher EDL thickness for the said value. Therefore, the magnitude of primary flow velocity is expected to be higher for lower $\alpha$, lower $\kappa$, and higher $d$, as witnessed in figure 5(a)-(c). Quite notably, with an increasing magnitude in the flow velocity for $\kappa = 12$, $d = 0.3$ and $\alpha = 1$, the magnitude of the Coriolis force $(\text{Re}_\omega \times \mathbf{u})$ increases since it linearly scales with the flow velocity. Thus, the higher Coriolis force for $\kappa = 12$, $d = 0.3$ and $\alpha = 1$ results in a noticeable depression in the geostrophic core as seen in 5(a)-(c).

However, as observed in figure 5(b), the difference between the magnitude of fluid velocities obtained for two consecutive values of $\kappa$ within its considered range $12 < \kappa < 50$ becomes minute. This observation is attributed to the close competition between the thicker influencing regime of smaller electroosmotic effect for the lower value of $\kappa = 12$ and the thinner influencing regime of higher electroosmotic force for the higher value of $\kappa = 50$. To support this argument, we have shown in figure 6(c) the variation of the electroosmotic force with a change in the DH parameter of EL. In figure 6(c), we can observe that despite having a higher magnitude of $\psi$ (as observed in figure 6(a)), the magnitude of the electroosmotic force is smaller for $\kappa = 12$. On the contrary, for $\kappa = 50$ (see figure 6(c)), the electroosmotic force becomes higher, albeit the magnitude of $\psi$ is seen to smaller (see figure 6(a)). In this variation, as per the definition of the electroosmotic force $(\kappa^2 \psi)$, the higher counter-ion density of the electrolyte for $\kappa = 50$ gives rise to the higher magnitude of the electroosmotic force. However, due to thinner EDL for $\kappa = 50$ as shown in figure 6(a), the influence of electroosmotic force being acted in the domain is comparatively smaller than that for $\kappa = 12$ (see figure 6(c)). In other words, the thicker influencing regime of the electroosmotic effect corresponding to the thicker EDLs for $\kappa = 12$ plays a dominant role here. On account of this, a minute variation in the electroosmotic force is obtained in figure 6(c) for a change in $\kappa$ from 12 to 50. Consequently, as observed in figure 5(b), the difference in the magnitude of $u$ velocity profile becomes small when $\kappa$ varies from 12 to 50. Precisely, it is because of this reason; the peak of $u$ velocity for both the values of $\kappa$ (12 and 50), as observed in figure 5(b), occurs approximately at the same $z$-location. Notably, this observation signifies a small change in the Ekman layer thickness with a change in $\kappa$.

Also, we observe in figure 5(a) that the velocity profile obtained for $\alpha = 20$ exhibits a small spike in the region close to the channel wall, albeit the appearance of such a spike is absent for the other cases under consideration. The presence of this spike in the velocity profile is attributed to the relatively larger Darcy drag force in the PEL region, realized for $\alpha = 20$. For other values of $\alpha$ (= 1 and 10), the electroosmotic force dominates the flow dynamics as the drag force is comparatively lower. The higher electroosmotic force wipes out the appearance of a small spike from the respective velocity profiles, as seen in figure 5(a). The influence of these parameters is also significant for the development of secondary flows in the present flow configuration. Note that the formation of secondary flow vortices is of paramount significance towards meeting the demand for essential fluidic functionalities of the current endeavour, i.e., the rapid and enhanced mixing of the species being transported through the channel. Paying attention to this aspect, we make an effort to critically review the development of secondary flows in greater detail in the forthcoming section.

### 6.2  *Secondary flow*
Development of secondary flows: Consistency with the experimental observations

The vortex structure pertaining to secondary flow obtained in the present study is exactly similar to the double-vortex configuration reported in the literature (Ng & Qi 2015). For the clear visualization of the double-vortex, we show in figure 7(a), the contour plot of stream function obtained for $\kappa_p = 10$, $d = 0.05$, $\alpha = 1$, $G = 0.5$, $\text{Re}_\omega = 100$ and $\kappa = 50$. As depicted in figure 7(a), the secondary flow assumes a double-vortex configuration with a clockwise vortex in the upper half $(z > 0)$, and a counter-clockwise vortex in the lower half $(z < 0)$ of the channel. It is worth to mention here that the geometrical features of these vortices are reminiscent of the structure of the geostrophic vortices as reported in the literature (Ng & Qi 2015; Pedlosky 1987). Also, the double-vortex configuration, as observed in figures 7(a), becomes similar with those captured in experimental investigations as well as numerical analysis (Kaushik *et al.* 2017b; Kheshgi & Scriven 1985; Masliyah 1980; Ng & Qi 2015; Speziale 1982), albeit different from the perspective of quantitative comparison.

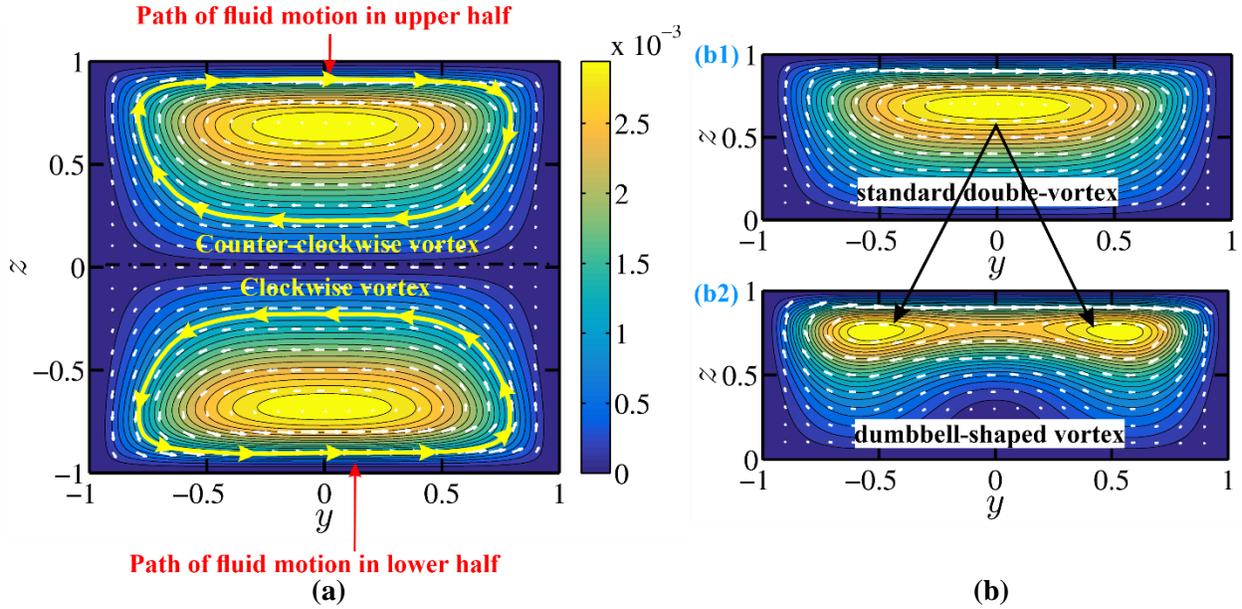

FIGURE 7 (color online): (a) Contour plot showing the double-vortex configuration, with a counter-clockwise vortex in the upper half of the cross-section and clockwise vortex in the lower half. The analysis is performed for $\kappa_p = 10$, $d = 0.05$, $\alpha = 1$, $G = 0.5$, $\text{Re}_\omega = 100$ and $\kappa = 50$. Note that for depicting the symmetric nature of the vortices in each half of the channel, we plot the absolute stream function in (a). (b) Contour plots of (b1) standard double-vortex and (b2) dumbbell-shaped vortex demonstrating the shift in the centroid of the vortex. The standard double-vortex in (b1) is obtained for $\kappa_p = 10$, $d = 0.05$, $\alpha = 20$, $G = 0.5$, $\text{Re}_\omega = 100$ and $\kappa = 50$ whereas the dumbbell-shaped vortex in (b2) is obtained for $\kappa_p = 10$, $d = 0.3$, $\alpha = 1$, $G = 0.5$, $\text{Re}_\omega = 100$ and $\kappa = 12$. Note that the values of the parameters considered for obtaining the standard double-vortex in (b1) are similar to that of (a).

Pertinent to this analysis of the secondary flow vortices, we make an attempt here to discuss the formation of different vortex configurations under the influence of PEL modulated electrostatics. In the rotational fluidics, the non-trivial geometrical features of the secondary flow such as the double-vortex configuration come into the picture because of the Coriolis force. The fluid under the influence of Coriolis force moves towards the vertical walls of the channel, from where it turns to the horizontal walls and then to the oppositely placed vertical walls of the channel (refer to the path of fluid motion shown in figure 7(a)). The path of such fluid motion in the entire cross-section gives rise to the formation of double-vortex configuration. Such a vortex structure includes the counter-clockwise vortex in the upper half and the clockwise vortex in the lower half as shown in figure 7(a).

*Formation of standard double-vortex configuration*

For a comparatively smaller strength of the Coriolis force, the motion of the fluid as discussed above (in the context of figure 7(a)) remains undisturbed. In effect, the weak secondary flow corresponding to the smaller strength of the Coriolis force makes a parallel arrangement of the streamlines at the middle section (geostrophic core) and the horizontal walls (Ekman layer) of the channel (see figure 7(a)-(b1)). Such vortex geometry is commonly observed in the paradigm of rotational microfluidics and defined as *the standard double-vortex* since its shape resembles what has been observed in the literature for both low Rossby number flows as well as low rotational Reynolds number flows (for weak Coriolis force) (Kheshgi & Scriven 1985; Nandakumar *et al.* 1991; Ng & Qi 2015). Note that such a small magnitude of Coriolis force $\text{Re}_\omega \times u$ is attributable to the weak strength of the primary flow (as discussed previously). As a result, we observe in figure 7(b1) that the higher PEL drag $(\alpha = 20)$ or higher DH parameter of EL $(\kappa = 50)$ or thin PEL $(d = 0.05)$, leads to the development of a standard double-vortex. Quite notably, combined effects of the weak primary flow strength (refer to figure 5(a)-(c) for primary flow) and the weak Coriolis force result in the standard double-vortex structure, as discussed above.

*Formation of dumbbell-shaped vortex*

For the stronger Coriolis force, the magnitude of secondary flow increases through the Ekman layer (Ng & Qi 2015). It may be mentioned here that the stronger Coriolis force arises due to the higher magnitude of the primary flow strength. However, in contrast to the confinement of the secondary flow only in the Ekman layer, the primary flow is present over the entire cross-section of the channel. This feature is visualized in figure 5(a)-(c) wherein the primary flow seems to be present in the entire cross-section. Whereas the secondary flow, as shown in the contour plots (obtained for comparatively higher Coriolis force) of figures 8 and 9, is found to be present near the horizontal walls of the channel only. This observation is suggestive of the fact that the secondary flow due to its weak magnitude in the geostrophic core is getting compressed by the primary flow therein. As a result of this phenomenon together with the symmetric nature of the primary flow about the center of the channel, the secondary flow gets compressed in all directions from the geostrophic core. The overall effect of this underlying phenomenon gives rise to the shifting of fluid mass towards the corners and results in the formation of the separate centroids in each quadrant of the channel cross-section (refer to figure 7(b2)). Note that the shifting of fluid mass also causes the compression of streamlines near the horizontal walls of the channel (refer to figure 7(b2)). Under such conditions, the shape of the vortex formed in each half of the cross-section (upper and lower) is reminiscent of the dumbbell shape, thus termed as the *dumbbell-shaped vortex* (Ng & Qi 2015). As per the definition of the Coriolis force, to obtain such a dumbbell-shaped vortex, the rotational system needs comparatively higher magnitude of fluid velocity. As established in the literature, such requirement of the higher velocity field in the rigid rotating microfluidic channel gets supplemented by the strong electroosmotic flow (Ng & Qi 2015). It is worth mentioning here that the strength of the electroosmotic flow can be further enhanced by the use of grafted polyelectrolyte layer on the walls of the fluidic channel. The thicker electrical double layer formed at the walls of the soft microfluidic channels provides stronger electroosmotic flow and assist in the building of the Ekman layer. On account of this advantageous feature of the soft microchannel, we observe the dumbbell-shaped vortex in figure 7(b2) resulting from the higher primary flow (refer to figure 5(a)-(c)) obtained for the lower PEL drag $(\alpha = 1)$, lower DH parameter of EL $(\kappa = 12)$ and higher PEL thickness $(d = 0.3)$.

Secondary flow structure: Influence of the Physico-chemical parameters

Next, we show, in figures 8(a)-(d) and 9(a)-(c), the contour plots of secondary flow, obtained for different parameters pertinent to this analysis. Note that in figure 8(a)-(d), the effect of drag parameter $\alpha$ (= 1 and 20) and DH parameter of EL $\kappa$ (= 12 and 50) on the secondary flow vortices are portrayed, while figures 9(a)-(c) describes the secondary flow structure for different PEL thickness $d$ (= 0.05, 0.1 and 0.3). In figures 8(a)-(d) and 9(a)-(c), we find the dumbbell-shaped vortices, except for a higher drag parameter $\alpha = 20$ and lower PEL thickness $d \leq 0.1$. The vortex in each half (upper and lower half) of the channel cross-section divides itself to form a pair of vortices, each resembling the shape of the dumbbell lobe. Note that the formation of these compressed stream tubes, which are the consequences of the combined effect of the higher electroosmotic force and the Coriolis force on the underlying transport, signifies the higher Ekman pumping in the channel. Thus, it may be mentioned here that the presence of these dumbbell-shaped vortices in the secondary flow structure is an indication of the presence of higher electroosmotic force as well as the resulting Coriolis force (since the strength of the Coriolis force being $Re_\omega \times \mathbf{u}$ indirectly depends on the electroosmotic force) in the present flow configuration. Therefore, the higher the strength of these forces, the higher will be the strength of dumbbell-shaped vortices appearing in the secondary flow structure.

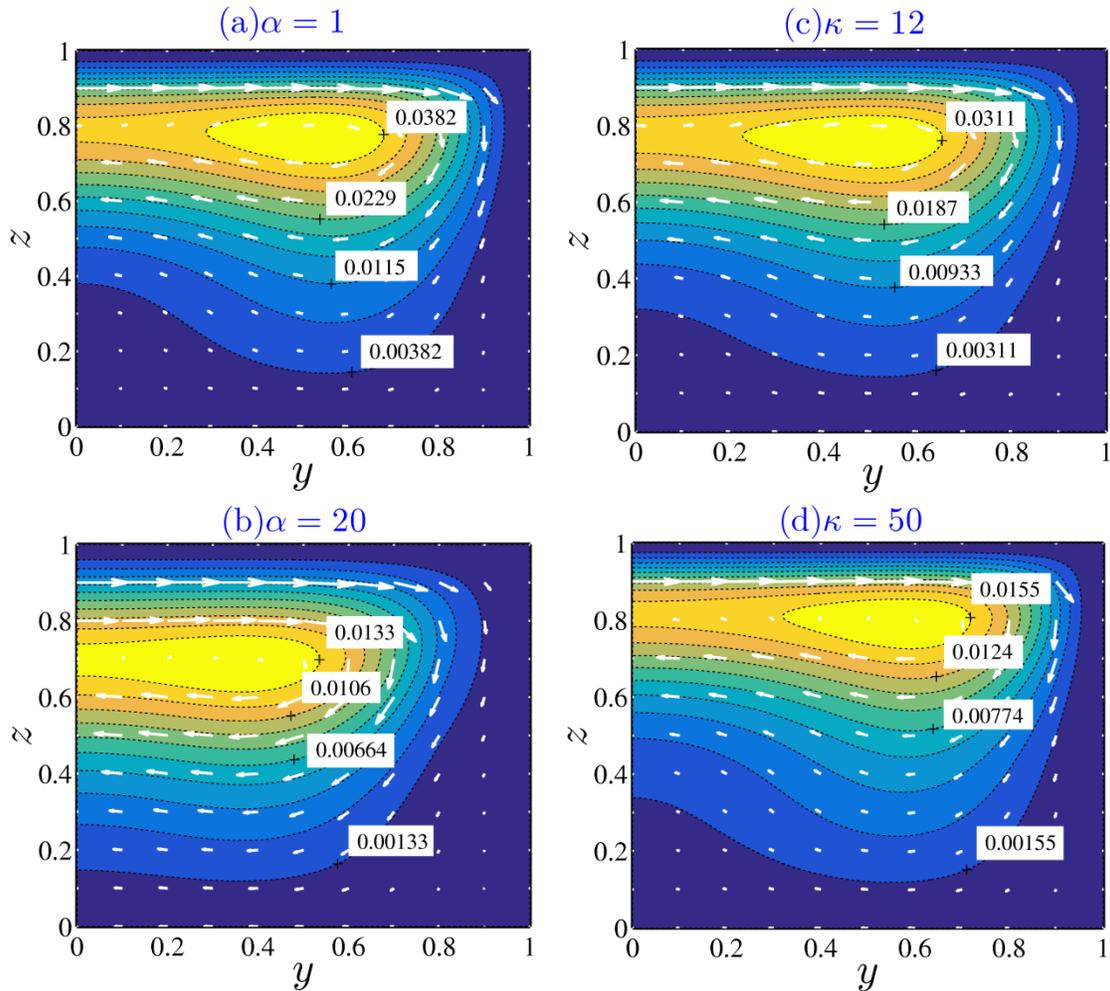

FIGURE 8 (color online). Contour plots show the secondary flow distribution in the channel for different values of (a)-(b) drag parameter $\alpha$ (= 1 and 20) and (c)-(d) DH parameter of EL $\kappa$ (=12 and 50). The other parameters considered for this analysis are: (a)-(b) $\kappa_p = 10$, $d = 0.2$, $G = 0.5$, $Re_\omega = 100$, $\kappa = 12$; (c)-(d) $\kappa_p = 10$, $d = 0.2$, $\alpha = 5$, $G = 0.5$, $Re_\omega = 100$.

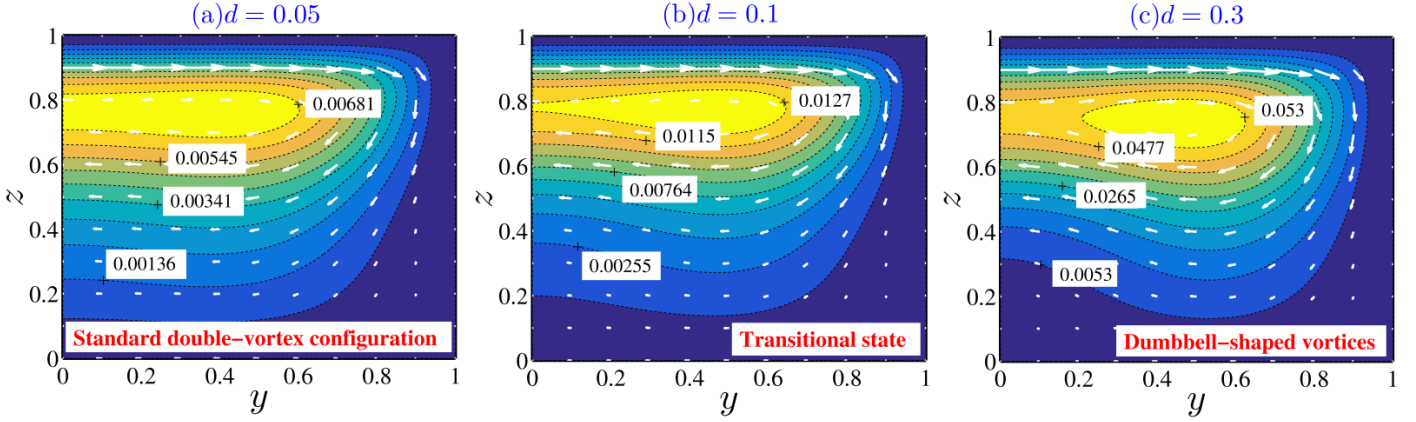

FIGURE 9 (color online): Contour plots show the secondary flow distribution in the channel for different values of PEL thickness $d$ (= 0.05, 0.1 and 0.3). The other parameters considered for this analysis are: $\kappa_p = 10$, $\alpha = 5$, $G = 0.5$, $\mathrm{Re}_\omega = 100$ and $\kappa = 12$.

*Influence of the drag parameter*

In figure 8(a)-(b), we observe the dumbbell-shaped vortices for $\alpha = 1$ and standard double-vortex configuration for $\alpha = 20$. In case of a higher drag parameter $\alpha (= 20)$, Darcy's frictional drag offered by the PEL dominates over the electroosmotic force and reduces the fluid velocity. Notably for $\alpha = 20$, we observe a small spike in the axial velocity profile in figure 5(a). Thus, following the definition of the Coriolis force ($\mathrm{Re}_\omega \times \mathbf{u}$), it can be argued here that the strength of the Coriolis force will be reduced for $\alpha = 20$. Precisely, this reduction in the strength of the Coriolis force reduces the Ekman pumping power, as manifested by the diffused streamline pattern near the horizontal walls, and results in the formation of the standard double-vortex configuration in figure 8(b). On the contrary, for $\alpha = 1$, a relatively smaller magnitude of the frictional drag force as compared to that of $\alpha = 20$ strengthens the PEL modulated electroosmotic flow near the walls. This results in the higher secondary flow velocities in the channel and a stronger Ekman pumping near the horizontal walls. Consequently, we observe the dumbbell-shaped vortices for $\alpha = 1$ in figure 8(a).

*Influence of the DH parameter of EL*

In figure 8(c)-(d), we observe an insignificant change in the shape of vortex when the value of $\kappa$ changes from 12 to 50. However, the value of stream function is observed to be high for $\kappa = 12$ than that of $\kappa = 50$, signifying the strength of higher secondary flow for $\kappa = 12$. This observation is attributed to the close competition between a thicker influencing regime of lower electroosmotic force for $\kappa = 12$ and a thinner influencing regime of higher electroosmotic force for $\kappa = 50$ (refer to figure 6(c) and section 6.1 for a detailed discussion). This complex variation in the electroosmotic force and its effective influencing regime near the walls (in figure 6(c)) with a change in $\kappa$ gives rise to a minimal change in the primary flow, as observed in figure 5(b). It is because of this reason the difference in the velocity profiles as well as its consequences to establish a difference in the thickness of the Ekman layer for the chosen values of $\kappa$ becomes very small, as witnessed in figure 5(b). Notably, this minute variation in the Ekman layer thickness with $\kappa$, as observed in figure 5(b), is further reflected by a small change in the vortex structure in figure 8(c)-(d). It may be mentioned here that the influence of the Coriolis force on the underlying transport gives rise to this correlative effect between the primary flow (in figure 5(b)) and the secondary flow (in figure 8(c)-(d)). Note that the relatively thicker influencing regime of the electroosmotic force for $\kappa = 12$ assists in the formation of the dumbbell-shaped vortices with comparatively higher magnitude, as seen in figure 8(c).

*Influence of PEL thickness*

Next, in figure 9(a)-(c), we show the structure of the secondary flow vortices developed in the flow field for three different values of $d = 0.05$, 0.1 and 0.3 respectively. We observe in figure 9(c) the existence of dumbbell-shaped vortices for higher PEL thickness $(d = 0.3)$, having higher strength of the Ekman pumping as well. On the other hand, for $d = 0.1$, the vortex structure is in the transition state between the standard double-vortex configuration and the dumbbell-shaped vortex, with a relatively lower strength of the Ekman pumping. Note that the special case of $d = 0.05$, which is almost closer to the case of a rigid channel, shows a perfect standard double-vortex configuration in figure 9(a). We attribute these observations to the variation of the electroosmotic force under the modulation of PEL thickness. The higher the PEL thickness, higher will be the electrostatic potential, as shown in figure 6(b), and this higher potential boosts up both the electroosmotic force (directly) as well as the Coriolis force (indirectly following the magnitude of flow velocity). Consequently, the higher magnitude of the Coriolis force arising from the relatively higher PEL thickness ($d = 0.3$) gives rise to higher Ekman pumping and hence results in the formation of the dumbbell-shaped vortices as apparent from figure 9(c). In contrast, the transitional feature of the secondary flow vortex for smaller PEL thickness ($d = 0.1$), as seen in figure 9(b), is due to the reduction in the strength of both the electroosmotic and Coriolis forces for the pertinent case.

Quite notably, the transitional state of vortex structure obtained for $d = 0.1$ is showing a similar qualitative configuration with the vortices developed in the rotational flows through a rigid, narrow fluidic channel. To substantiate this inference further in the context of the present analysis, we show in figure 9(a) the vortex structure for PEL thickness $d = 0.05$. As can be seen in figure 9(a), we can get a perfect standard double-vortex pattern for $d = 0.05$, which is similar to the reported vortex structure formed in a rigid rotating narrow fluidic channel (Ng & Qi 2015). Note that for such a small PEL thickness $d = 0.05$, the influence of PEL on the flow dynamics can be considered to be negligible, thus the underlying transport feature signifying the flow scenario of a rigid, narrow fluidic channel with a common pressure-driven configuration.

The secondary flow vortices being developed in the flow field of the present configuration are essential for the mixing phenomenon in the rotating channel. It is worth to mention here that the augmented secondary flow strength for the cases of $\kappa = 12$ (figure 8(c)) and $d = 0.3$ (figure 9(c)) alongside the formation of standard double-vortex configuration observed for $\alpha = 20$ in figure 8(b), plays an essential role in the mixing dynamics when the rotational speed of the channel is considerably high. From the ongoing discussion, it can be inferred that the flow dynamics in a rotating microfluidic channel with a built-in polyelectrolyte layer exhibits the modulation of complex secondary flow vortices, as evident from figures 8 and 9. The attributable feature behind this observation is the grafted polyelectrolyte layer modulated enriched interfacial electrostatics and its effect on the alteration in the magnitude of the rotation induced forcing. To be precise, as a result of this analysis presented till now, we believe that the imposition of soft polyelectrolyte layers offers an avenue in tuning the structure as well as the strength of recirculation zones (equivalently, the secondary flow vortices) being formed in the flow field. In view of this unique feature of the soft polyelectrolyte layer on the electrohydrodynamics in the rotating microfluidic channel, we next make an effort to look at its extended effect on the underlying mixing dynamics critically.

## 7 Effect of PEL modulated electrohydrodynamics on the mixing: analysis, results, and discussion

In line with the prime objective of the present endeavour, we investigate in this section the mixing dynamics in the soft microchannel embedded in the rotating platform. We look at the underlying mixing from the perspective of both qualitative and quantitative assessments. To perform the qualitative analysis, we focus on the Poincaré maps at different axial locations of the channel. On the other hand, to quantify the mixing strength, we analyze the variation in the mixing index obtained by using the concept of entropy of mixing. Both these methods are based on the Lagrangian approach, in which the passive tracer particles of multispecies are traced in the 3D computational domain with the help of Eulerian velocity fields. We now systematically discuss the procedure and results obtained from these methods as follows.

### 7.1  *Poincaré maps*

To obtain the qualitative information of the mixing dynamics, we focus on the Poincaré maps obtained at different axial locations of the channel. The traces of the passive tracer particles in 3D computational domain produce the Poincaré maps. These traces can be calculated by using the following Lagrangian equations, which rely on the Eulerian velocity fields $(u,v,w)$ obtained from the previous calculations (section 2 and 3).

$$u = \frac{\partial x}{\partial t} \tag{7.1.a}$$

$$v = \frac{\partial y}{\partial t} \tag{7.1.b}$$

$$w = \frac{\partial z}{\partial t} \tag{7.1.c}$$

It can be seen that the equations (7.1) are integrable with respect to time. In contrast, the Poincaré maps show the $y$ and $z$ locations of the particles in the cross-section of the channel at the desired axial (along $x$-coordinate for this analysis) location. This implies that the secondary flow velocities should be integrated with respect to the axial coordinate $x$. In doing so, we divide equations (7.1.b) and (7.1.c) by (7.1.a).

$$\frac{\partial y}{\partial x} = \frac{v}{u} \tag{7.2.a}$$

$$\frac{\partial z}{\partial x} = \frac{w}{u} \tag{7.2.b}$$

The magnitudes of the velocities $u$, $v$ and $w$ at the updated locations ($y$ and $z$) are obtained by using the function-griddedInterpolant(cubic) in Matlab 2019a. Unless mentioned specifically, the step size in the axial direction, i.e., $\Delta x$ is taken as $10^{-3}$, and the total length of the channel, to be consistent with our previous assumption made in Section 4, is considered as $10^3$. In figure 10(a), the sample trajectories of a few tracer particles of two different types [particles $P_{a,1}$ and $P_{a,2}$ in figures 10(a) and 10(b)] are shown in the 3D computational domain.

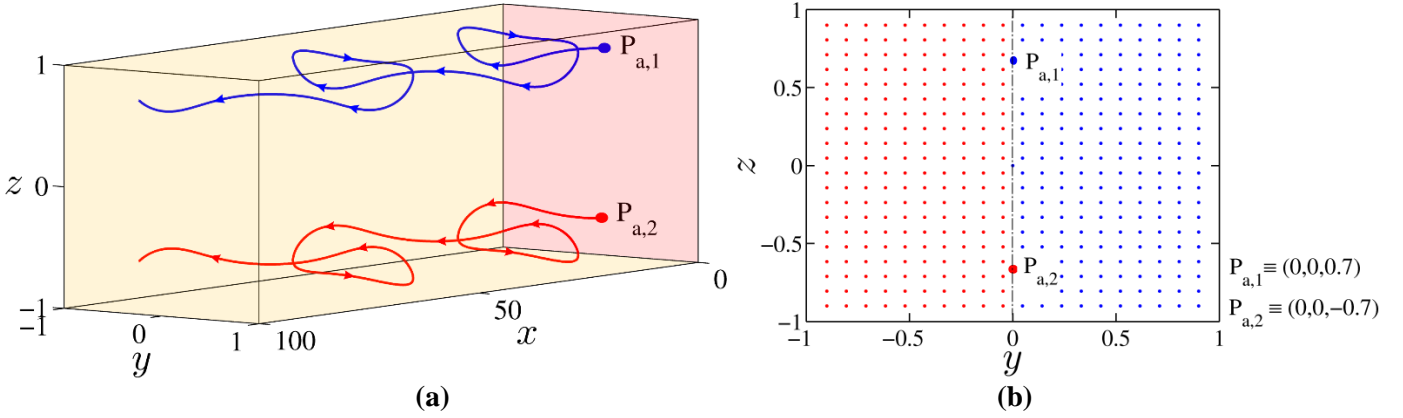

**FIGURE 10 (color online). Sample trajectories of tracer particles in the 3D computational domain:** Trajectories of a few tracer particles ($P_{a,1}$ and $P_{a,2}$) in a 3D computational domain are shown in (a) for $\kappa_p = 10$, $d = 0.3$, $G = 0.5$, $Re_\omega = 150$, $\kappa = 12$ and $\alpha = 5$. $P_{a,1}$ and $P_{a,2}$ are initially located at (0,0,0.7) and (0,0,-0.7) respectively. **Poincaré map at inlet:** Specified initial positions of $20 \times 20$ tracer particles in the $y - z$ cross-section shown in (b). The colors of the particles: red and blue, distinguish the species considered for the calculation.

In figure 10(a), we observe that the tracer particles ($P_{a,1}$ and $P_{a,2}$), trapped in the dumbbell-shaped vortices (depicted in figures 8 and 9), follow a *periodic helical* motion, primarily attributed to the periodic motion of the fluid particles therein. Also, the trajectory of a blue particle $P_{a,1}(\equiv(0,0,0.7))$ released from the upper half of the channel exactly forms a mirror image of the trajectory of red particle $P_{a,2}(\equiv(0,0,-0.7))$ released from the lower half. Note that the pathlines of particles $P_{a,1}$ and $P_{a,2}$ depicted in figure 10(a) show similar morphology with the corresponding streamline contours shown in figures 8 and 9. By depicting the traces of particles $P_{a,1}$ and $P_{a,2}$ in figure 10(a), we intend to convey that in the present flow configuration, the stream of the particles (represented by blue particle $P_{a,1}$) released from the upper half of the channel (about the line $z = 0$ and parallel to $y$-axis) will mix insignificantly with the particles (represented by red particle $P_{a,2}$) being released in the lower half. Considering this aspect, we inject the particles of the different species through the left half and right half (two symmetry halves about the line $y = 0$ and parallel to $z$-axis) of the channel, as shown in figure 10(b). The colors of the particles, red for the right half and blue for the left half (if observed from the inlet), distinguish the species considered for the calculation.

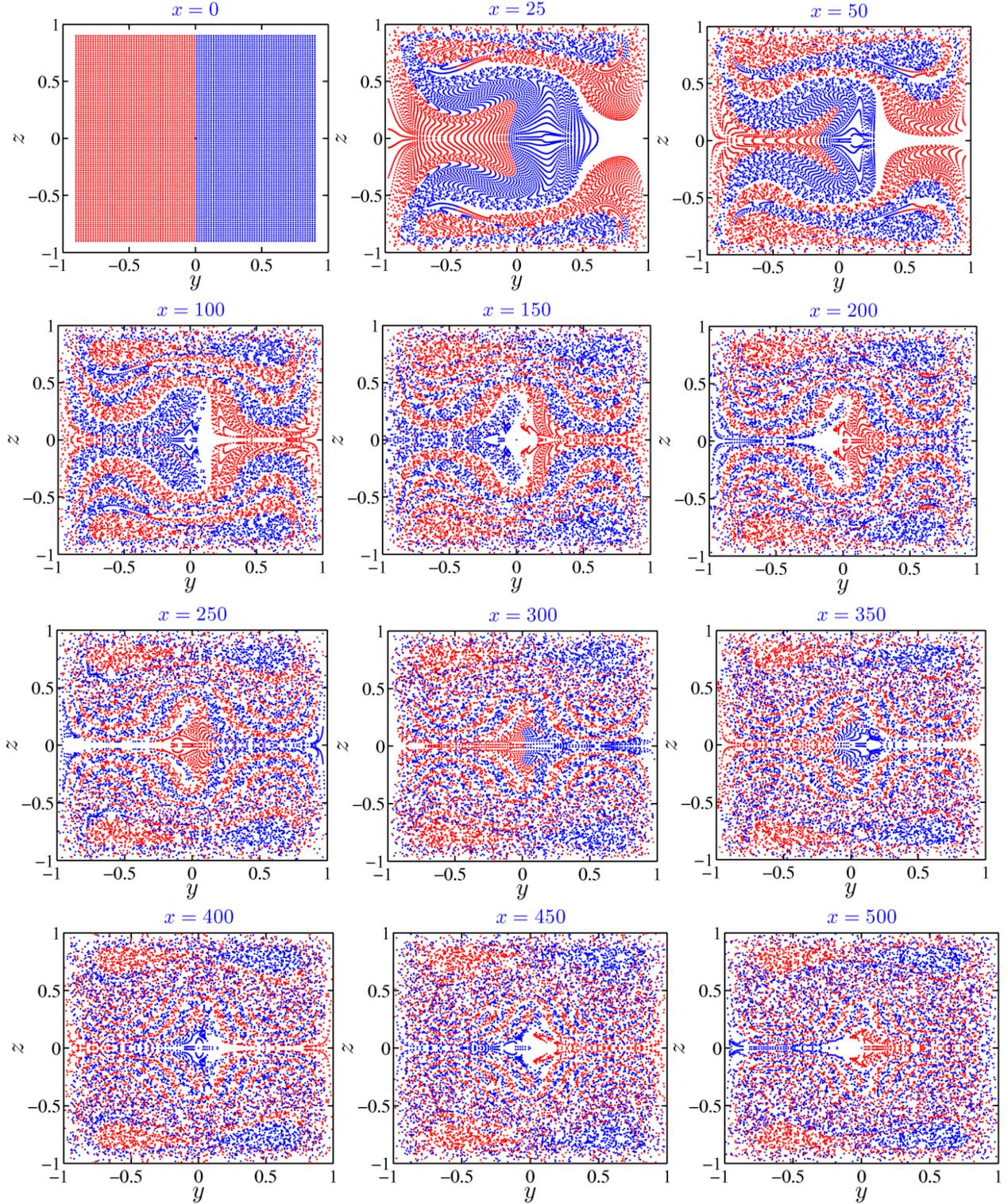

FIGURE 11 (color online). **Axial evolution of the Poincaré maps:** Plots show the axial evolution of the Poincaré maps, obtained for $\kappa_p = 10, d = 0.3, G = 0.5, \text{Re}_\omega = 150, \kappa = 12$ and $\alpha = 5$. These parameters are similar to that of considered for figure 4. The plots are obtained at: $x = 0, 25, 50, 100, 150, 200, 250, 300, 350, 400, 450$ and $500$. The number of particles considered is 100x100.

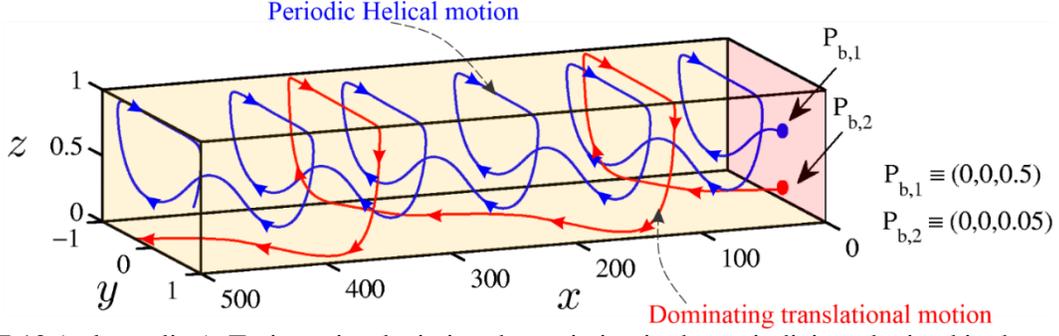

FIGURE 12 (color online). Trajectories depicting the variation in the periodicity, obtained in the motion of two different particles ($P_{b,1}$ and $P_{b,2}$) initially located at (0,0,0.5) and (0,0,0.05), respectively. Note that the plots are depicted in the upper half section of the channel. The other parameters considered for this analysis are: $\kappa_p = 10$, $d = 0.3$, $G = 0.5$, $\mathrm{Re}_\omega = 150$, $\kappa = 12$ and $\alpha = 5$.

Having outlined the solution procedure at the beginning of this section, we now discuss the axial evolution of the Poincaré maps depicted in figure 11 for $\kappa_p = 10$, $d = 0.2$, $G = 0.5$, $\mathrm{Re}_\omega = 150$, $\kappa = 12$, and $\alpha = 5$. We observe in figure 11 that, beyond the location $x = 350$, the change in the Poincaré maps is insignificant except at the geostrophic core of the channel. This observation is suggestive of the efficient mixing of species, albeit qualitative, beyond $x = 350$ and except the core region. As shown in figure 12, the attributable factor behind this observed discrepancy at the core region is the delay in attaining the periodic behaviour of the fluid particles therein (precisely, the periodic motion is getting delayed). The location of the blue particle $P_{b,1}(\equiv(0,0,0.5))$ and the red particle $P_{b,2}(\equiv 0,0,0.05)$ shown in figure 12, represents the particles being released from the non-geostrophic and geostrophic region of the channel, respectively. Each belongs to different species (red/blue). In the non-geostrophic region, the periodicity in the motion is quickly obtained almost after $x = 50$, visualized from the trajectory of particles in the dumbbell-shape for $x > 50$ (see figures 11 and 12(blue particle $P_{b,1}$)). However, the trajectories of the particles at the core region (red particle $P_{b,2}$ in figure 12) attain well-defined periodicity at a sufficiently larger distance from the inlet ($x = 500$), as observed in figures 11 and 12 (see for red particle $P_{b,2}$). The periodicity of this motion depends on the strength of secondary flow developed in the flow field. However, due to the very small magnitude of $v$ and $w$ velocities at the geostrophic core (as shown by the surface plots in figure 4(b) and 4(c)), the strength of secondary flow and its effect in developing the periodic fluid motion becomes minuscule in the interior. Consequently, the primary flow ($u$-velocity) imposes the dominating translational motion instead of a periodic motion to the fluid particles at the geostrophic core, as observed in figure 12. This flow feature results in qualitatively poor mixing at the geostrophic core as compared to that at the non-geostrophic region. This observation also implies that for the given set of governing parameters, the particles mostly passing through the core region encounter a relatively larger mixing length[3]. For example, to achieve approximately complete mixing, the particles (red particle $P_{b,2}$) being released from the core region are expected to travel a considerable distance in the channel. This aspect is visualized from the depicted variations in both the figures 11 and 12: in figure 11, the stream of particular species at the core region remains poorly mixed even up to axial distance $x = 500$; and in figure 12, the time taken by red particle $P_{b,2}$ to complete the particular number of circulations (say, two in this case) is higher than that of a blue particle $P_{b,1}$.

Influence of Physico-chemical parameters of PEL on mixing

Next, in figure 13, we show the influence of the physicochemical parameters of the grafted PEL on the mixing performance using the Poincaré maps. The other parameters considered to obtain the current plots (mentioned in the caption of figure 13) are the same as that considered for depicting the secondary flow in figures 8 and 9. Note that in figure 13, we intentionally show the Poincaré maps at a section corresponding to the axial location $x = 500$, where the candidate streams of the species are partially mixed for all the cases considered. At a considerably distant axial location, precisely, at the location closer to the outlet of the channel, a change in the magnitude of pertinent parameter does not bring a noticeable difference in the Poincaré maps. This observation justifies a complete mixing at that location.

In figure 13, we observe that the mixing quality obtained for $\alpha = 20$, $\kappa = 12$ and $d = 0.3$ at an axial location $x = 500$ are qualitatively higher than that of their respective counterpart. Also, as seen in figure 13, for the other values of the chosen parameters, the particles of different species are yet to be significantly mixed until the location $x = 500$ is reached. This observation underlines the competitive strength of the dumbbell-shaped vortices vis-à-vis the effect of standard double-vortex configuration on the underlying mixing phenomenon. For $\alpha = 20$, the size of the recirculation zone (standard double-vortex configuration) is larger, as seen in figure 8(b), whereas the strength of the dumbbell-shaped vortices for $\kappa = 12$ and $d = 0.3$, as seen in figure 8(c) and 9(c) respectively, becomes higher. This higher strength and size of the vortices obtained for the cases mentioned above further lead to an enhancement in the characteristics of the chaotic region thus resulting in enhanced mixing performance. This observation also infers that the mixing length, i.e., the distance from the inlet of the channel where the streams significantly mix, is comparatively smaller for the cases of $\alpha = 20$, $\kappa = 12$ and $d = 0.3$.

---

[3] Here, the mixing length referes to the length where substantial mixing occurs.

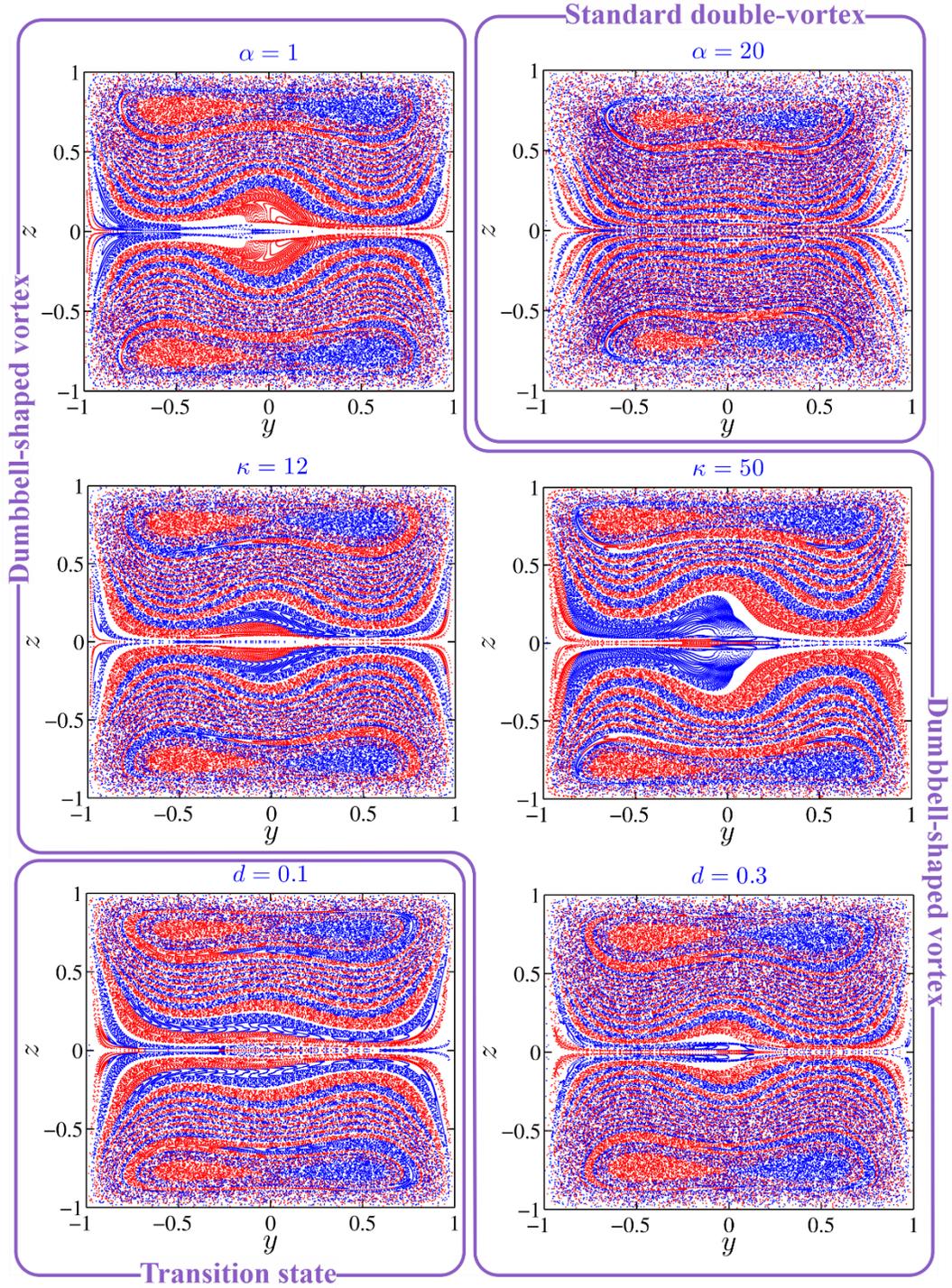

FIGURE 13 (color online). **Qualitative assessment for the influence of governing parameters on mixing**

**dynamics:** Poincaré sections obtained at $x = 500$ to show the influence of governing parameters on the underlying mixing dynamics. The parameters considered are: drag parameter $\alpha$ (= 1, 20); DH parameter of EL $\kappa$ (= 12, 50); and PEL thickness $d$ (= 0.1, 0.3). The other parameters considered for the calculation are: for variation in $\alpha$ - $\kappa_p = 10, G = 0.5, \text{Re}_\omega = 100, d = 0.2, \kappa = 12$; for variation in $\kappa$ - $\kappa_p = 10, \alpha = 5, G = 0.5, \text{Re}_\omega = 100, d = 0.2$; and for variation in $d$ - $\kappa_p = 10, \alpha = 5, G = 0.5, \text{Re}_\omega = 100, \kappa = 12$. The number of particles considered is 250 x 250.

In the context of this discussion, it is important to mention here that in figure 13, we observe a minute difference between the Poincaré maps of $d = 0.1$ and $d = 0.3$. Note that for $d = 0.1$, the present fluidic setup approaches toward resembling a rigid, narrow fluidic channel and, therefore, gives rise to the formation of secondary flow vortex, which is in a transition state between the standard double-vortex configuration and dumbbell-shaped vortex (discussed before in section 6.2). Notably, the area of the recirculation zone for this transition state at $d = 0.1$ is higher than that of the dumbbell-shaped vortices obtained for $d = 0.3$ (refer to figure 9(b)-(c)). In contrast, the strength of the vortices for $d = 0.3$ is higher than that of $d = 0.1$ as seen in figure 9(b)-(c). On account of this competition between the strength and the size of the vortices, we obtain almost an equal mixing performance for both the values of $d$ (=0.1, 0.3), as witnessed in figure 13. Also, it can be added here that the mixing performance obtained in a rigid channel can be higher than that of the channel having thicker PELs grafted on its wall. The vortex structure obtained for the case of minimal PEL thickness (say $d = 0.05$) has a very higher recirculation area (refer to figure 9(a)) as compared to that of the higher PEL thickness (cf. figure 9(c)). However, a smaller PEL thickness ($d = 0.05$) has some adverse consequences on the underlying mixing, as discussed from the quantitative perspective in the next Section 7.2. Also, in section 7.2, we will highlight the effect of a higher frictional drag, mainly stemming from the grafted PEL, on the mixing performance since it has similar adverse consequences.

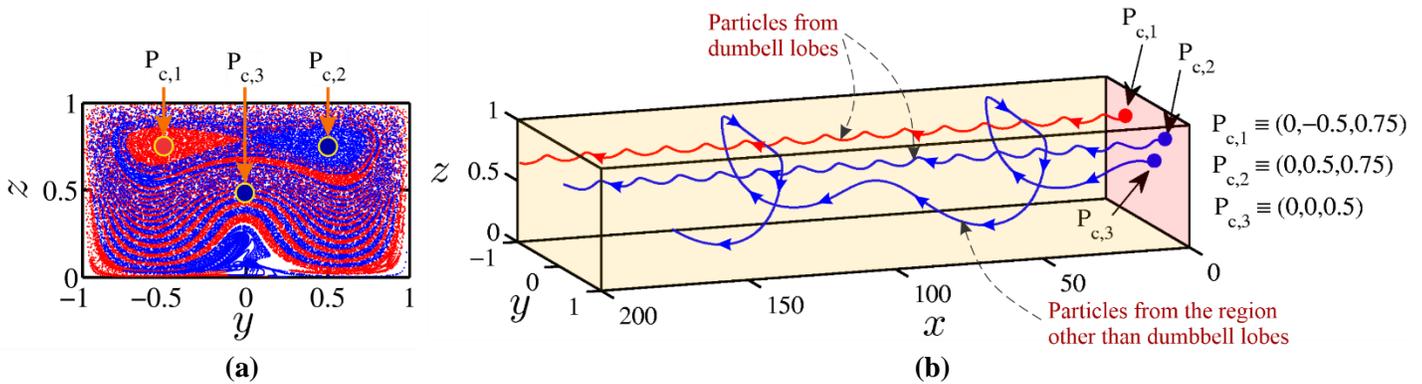

FIGURE 14 (color online). (a) Locations of the particles at x = 500: $P_{c,1}(\equiv(0,-0.5,0.75))$ and $P_{c,2}(\equiv(0,0.5,0.75))$ are located in the dumbbell lobes whereas $P_{c,3}(\equiv(0,0,0.5))$ is located in the remaining region of the dumbbell-shaped vortices. (b) Trajectories depicting the isolated motion of the particles $P_{c,1}(\equiv(0,-0.5,0.75))$ and $P_{c,2}(\equiv(0,0.5,0.75))$, from the particle $P_{c,3}(\equiv(0,0,0.5))$. The other parameters considered for this analysis are: $\kappa_p = 10$, $d = 0.3$, $G = 0.5$, $\text{Re}_\omega = 150$, $\kappa = 12$ and $\alpha = 5$. Note that for this analysis, we have considered the intermediate segment of the channel.

Also, we observe in figure 13 that the lobes of the dumbbell-shaped vortices trap the particles of both the species. This phenomenon adversely affects the mixing performance following the event discussed as follows. The particles from these lobes do not enter into the chaotic regime; instead, keep on moving following the undisturbed trajectories. Such trajectories are shown in figure 14(a)-(b). In figure 14(a)-(b), the plots demarcate the isolated motion of the particles $P_{c,1}(\equiv(0,-0.5,0.75))$ and $P_{c,2}(\equiv(0,0.5,0.75))$ located inside the dumbbell lobes, from a particle $P_{c,3}(\equiv(0,0,0.5))$ moving through the remaining region of the dumbbell-shaped vortex. It can be seen in figure 14 that the particles $P_{c,1}$ and $P_{c,2}$ from the dumbbell lobes never interact with a particle $P_{c,3}$ from the remaining dumbbell-shaped region. Throughout the length of the channel, the number density of particular species residing inside the dumbbell lobes (represented by particles $P_{c,1}$ and $P_{c,2}$) remains unaltered. As a result, the entropy of such discrete sections (lobes) remains constant, attributed primarily to non-intrusion of other species' particles into the lobes. This is a disadvantage of the rotational micromixers, even in the presence of the PEL modulated electrostatics. However, for $d = 0.1$ and $\alpha = 20$, the density of trapped particles inside the lobes is negligible due to the standard double-vortex configuration (refer to figure 13).

The insights gained from figure 13 will be of help for the design of the proposed rotating microfluidic devices, ensuring efficient mixing. To this end, the parameters modulating the PEL electrostatics $(\kappa, d)$, and the underlying hydrodynamics $(\alpha)$ seem to have huge consequences on the underlying mixing within the limit of asymptotic geostrophic plug flow. Having discussed the qualitative analysis of the mixing, we now focus on the quantification of the mixing dynamics in the following section.

### 7.2 *The entropy of mixing and mixing index*

To quantify the mixing dynamics following the Lagrangian approach, we analyze the variation in the mixing index, obtained by using the concept of entropy of mixing. In this method, the computational domain represented by the Poincaré sections in figure 13 is equally divided into a finite number of discrete cells. A task integrated with this approach is to estimate the probability of finding the particles of each species indexed to each cell. The entropy of mixing in that cell is the logarithmic function of this probability, and it is the total entropy of the considered Poincaré section if averaged over the entire map. The total entropy of mixing for each Poincaré section is given by,

$$S = -\sum_{i=1}^{N_C} \left[ \beta_i \sum_{j=1}^{N_S} \left( n_{i,j} \log\left(n_{i,j}\right) \right) \right] \quad (7.3)$$

In equation (7.3), $N_C$ is the number of flow cells, $N_S$ is the number of species, $i$, $j$ are the indices for cells, species, respectively, and $n_{i,j}$ is the fraction of species $j$ in the $i^{\text{th}}$ cell. Note that $\beta_i$ is the weighting function for each cell. It is zero if only the single type of species is present in the cell $i$, and it is unity if all kinds of species are present in the $i^{\text{th}}$ cell. Further, we normalize this local entropy on the scale of [0, 1] using the maximum entropy of that Poincaré section. The maximum entropy of the specific Poincaré section is given by,

$$S_{\max} = -N_C \log\left(\frac{1}{N_S}\right) \quad (7.4)$$

The normalized entropy (or mixing index $I_m$) is given by,

$$I_m = \frac{S}{S_{\max}} \quad (7.5)$$

The mixing index $I_m$ varies between '0' and '1', signifying '0' as the poor mixing quality and '1' as the complete mixing quality. Note that the value of the mixing index varies with the number of flow cells. The lower number of flow cells artificially increases the value of mixing index and vicé versa. Therefore, similar to the grid dependency test typically considered in the paradigm of computational fluid dynamics, the optimum number of flow cells can be finalized after performing the cell dependency test. However, the estimation of the mixing index using a higher number of cells increases the computational time for each Poincaré section. In view of that, we finalize $20 \times 20$ flow cells per Poincaré section to estimate the mixing index after performing the cell dependency test.

The axial variation of the mixing index under the influence of different governing parameters is shown in figure 15. In figure 15, we consider different values of (a) drag parameter of PEL $\alpha$ (= 1, 10, 20), (b) DH parameter of EL $\kappa$ (=12, 20, 50) and (c) PEL thickness $d$ (= 0.05, 0.1, 0.2 and 0.3). We observe from figure 15 that for all the parameters considered, the mixing index increases in the axial direction. This observation signifies the enhancement in the mixing performance with an increment in the distance from the inlet. The appearance of small oscillations in the plots of figure 15 is attributed to the discrepancy in the probability of finding particular species in a fixed cell at $x^{th}$ and $(x+1)^{th}$ Poincaré sections. The continuous motion of the particles in the 3D computational domain tunes this probability and, at times, adjusts the magnitude of the weighting function $\beta_i (0/1)$ of a particular $i^{th}$ cell between each consecutive iteration in the $x$ direction. As a consequence, we observe the appearance of oscillation in the mixing index variation along the axial direction of the channel in figure 15. Noteworthily, a continuous increment in the mixing index along the length of the channel indicates the efficient performance of the proposed microfluidic mixer. The influence of different other parameters on the mixing index is discussed in detail as follows.

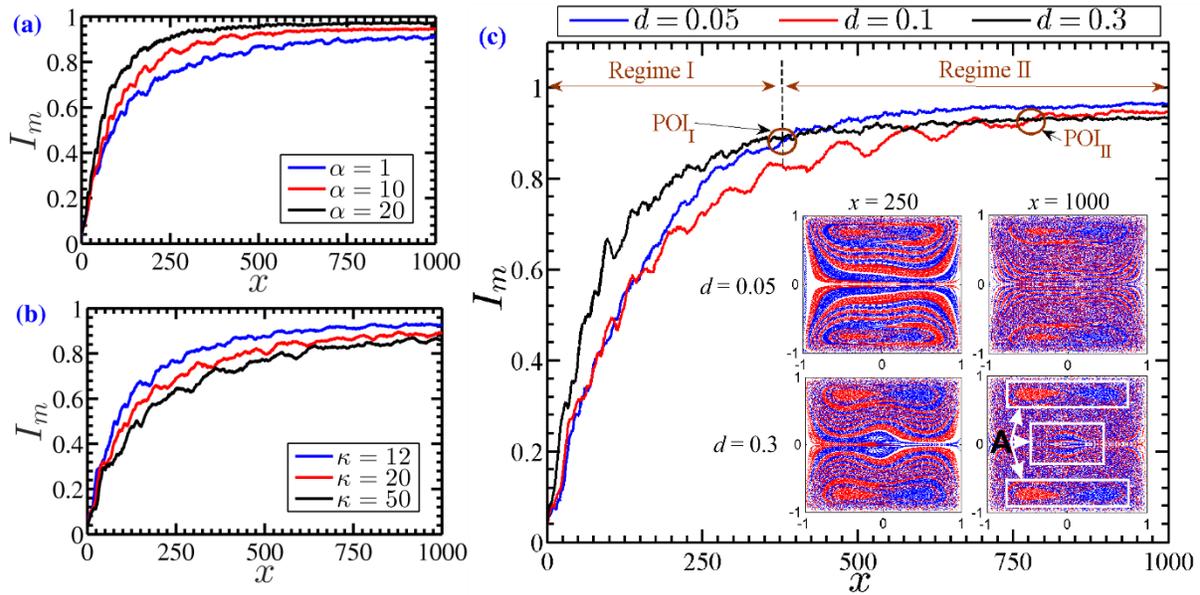

FIGURE 15 (color online). **Quantitative assessment for the influence of governing parameters on mixing dynamics:** Plots showing the variation of mixing index $I_m$ in axial direction for different values of (a) drag parameter of PEL $\alpha$ (= 1, 10 and 20), (b) DH parameter of EL $\kappa$ (= 12, 20 and 50), and (c) PEL thickness $d$ (= 0.05, 0.1, and 0.3). The other parameters considered for analysis are: (a) $\kappa_p = 10$, $G = 0.5$, $Re_\omega = 100$, $d = 0.2$, $\kappa = 12$; (b) $\kappa_p = 10$, $\alpha = 5$, $G = 0.5$, $Re_\omega = 100$, $d = 0.2$; and (c) $\kappa_p = 10$, $\alpha = 5$, $G = 0.5$, $Re_\omega = 100$, $\kappa = 12$. The insets in (c) show the Poincaré maps for both values of PEL thickness $d$(=0.05, 0.3) at two different axial locations $x = 250$ and $1000$.

In figure 15(a) and 15(b), we observe that the magnitude of the mixing index $I_m$ obtained for higher drag parameter $\alpha(=20)$ and lower DH parameter of EL $\kappa(=12)$, respectively, is higher than that of their respective counterparts. These observations are attributed to the higher area of standard double-vortex configuration for $\alpha = 20$ observed in figure 8(b), as well as the higher strength of the dumbbell-shaped vortices for $\kappa = 12$ observed in figure 8(c).

The plots in figure 15(c) show the variation of the mixing index $I_m$ with PEL thickness $d$ (= 0.05, 0.1, and 0.3). The insets in figure 15(c) show the Poincaré maps for both the values of PEL thickness $d\left(=0.05, 0.3\right)$ obtained at different axial locations $x$ = 250 and 1000. A soft microfluidic channel with $d = 0.05$ mimics a rigid narrow fluidic channel. In figure 15(c), we observe that, after a certain $x$ location labelled as the point of inflexion (POI$_I$) with regime I and II in figure 15(c), the mixing index for minimal PEL thickness $d = 0.05$ becomes higher than that of higher PEL thickness $d = 0.3$. Notably, this observation is in agreement with our prediction of mixing, as discussed in Section 7.1. In the comparison of $d$ =0.1 and 0.3, such point of inflexion, named as POI$_{II}$, appears at a comparatively larger distance from the inlet of the channel. It is very fact that the competition between the relatively larger area of the recirculation zone obtained for the smaller PEL thickness ($d = 0.05$ in figure 9(a) and $d = 0.1$ in figure 9(b)), and the higher strength of the dumbbell-shaped vortices for higher PEL thickness ($d = 0.3$ for figure 9(c)) plays a key role here. To discuss this aspect, we take the help of insets depicted in figure 15(c). As shown in the insets of figure 15(c), for $d = 0.05$, the size of the recirculation zone is higher than that of $d = 0.3$. This observation is applicable for both the axial locations $x = 250$ (in regime I) and $x = 1000$ (in regime II). Despite that, as observed in figure 15(c), the higher strength of the dumbbell-shaped vortices obtained for $d = 0.3$ dominating over the initial channel length, as shown by regime I. On the other hand, in the remaining length of the channel, identified by regime II, the higher area of the standard double-vortex obtained for $d = 0.05$ still prevails. Consequently, the value of $I_m$ is higher for $d = 0.3$ in regime I and $d = 0.05$ in regime II. Note that the mixing comparison obtained for $d = 0.1$ and $d = 0.3$ can be explained from a similar perspective, as discussed above. For this case, a closer look at the point of inflexion POI$_{II}$ is necessary.

Moreover, in consistence with the observations of figures 12 and 14, we observe a similar phenomenon in the insets of figure 15(c) as follows: the large isolated volume of species trapped in the dumbbell lobes and the geostrophic core of the channel for the case of $d = 0.3$ causes a reduction in the mixing index. This isolated region of species ('A') is shown in figure 15(c) for $d = 0.3$ and $x = 1000$. Notably, the region where the species reside as isolated or nearly isolated ('A') is minimal in case of $d = 0.05$ (refer to inset corresponding to $d = 0.05$ and $x = 1000$ in figure 15(c)), causing an increase in the mixing index in regime II of figure 15(c). It is also worth mentioning here that the observation obtained in figure 15(c) is not consistent with the drag parameter variation. Over the total length of the channel, the strength of the vortices for the drag parameter $\alpha = 20$ (see figure 8(b)) is comparatively high to show a very high mixing performance as compared to the case $\alpha = 1$ (figure 8(a)).

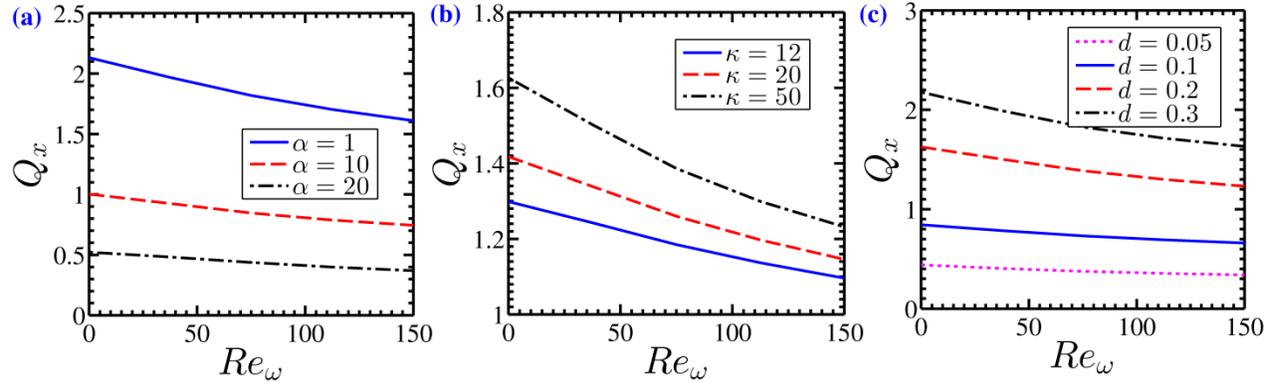

FIGURE 16 (color online): Plots showing the variation of axial flow rate $(Q_x)$ with a rotational speed of the channel $(\text{Re}_\omega)$ for different values of (a) drag parameter of PEL $\alpha$ (= 1, 10 and 20), (b) DH parameter of EL $\kappa$ (= 12, 20 and 50), and (c) PEL thickness $d$ (= 0.05, 0.1, 0.2, 0.3). The other parameters considered to plots these results are: (a) $\kappa_p = 10$, $G = 0.5$, $\text{Re}_\omega = 100$, $d = 0.2$, $\kappa = 12$; (b) $\kappa_p = 10$, $\alpha = 5$, $G = 0.5$, $\text{Re}_\omega = 100$, $d = 0.2$; and (c) $\kappa_p = 10$, $\alpha = 5$, $G = 0.5$, $\text{Re}_\omega = 100$, $\kappa = 12$.

In line with our observation and subsequent discussion in section 7.1, here in figure 15, it is observed that, for the fixed $\text{Re}_\Omega (=100)$, the mixing length obtained for higher $\alpha (=20)$, lower $\kappa (=12)$ and higher $d (=0.3)$, is comparatively smaller as compared to that of their respective counterparts. This observation constitutes the proof of the enhancement in the mixing efficiency under the influence of PEL modulated electrohydrodynamics in the rotating microfluidic platform. In this context, we would like to discuss an important point as follows: although the mixing efficiency obtained at the outlet, as depicted in figure 15, is relatively higher for higher drag parameter $(\alpha = 20)$ and very small PEL thickness $(d = 0.05)$ (a representing case of the insignificant effect of PEL, i.e., almost a rigid narrow fluidic channel), the net throughput, which is proportional to the axial velocity, becomes lesser for $\alpha = 20$ and $d = 0.05$, as shown in figure 16(a) and 16(c) respectively. Figure 16 depicts the variation of the axial flow rate $(Q_x)$ with the rotational speed of the channel $(\text{Re}_\omega)$ for different parameters mentioned in the caption. From the ongoing discussion, we would like to iterate here that achieving a higher mixing efficiency through the imposition of the grafted polyelectrolyte layer demands a compromisation in the net throughput for a relatively higher drag parameter $(\alpha = 20)$ and very small PEL thickness $(d = 0.05)$. As a result, for these relevant cases, the present soft narrow fluidic configuration will again imitate the similar characteristics of the rigid narrow fluidic channel, thus rendering an unacceptable outcome. However, for lower $\kappa(=12)$, the higher electroosmotic effect offers higher mixing performance as well as enhances the net axial throughput in the channel (see figure 16(b)). Thus, taking a note on this aspect, the judicial selection of the physicochemical structure of PEL seems to be imperative for the optimum mixing performance without negotiating the throughput in the proposed mixer substantially.

## 8  Summary

We theoretically investigate the flow dynamics and its consequences to the underlying chaotic mixing phenomenon in a soft rotating microchannel. We employ the variational calculus method for the flow field analysis, while we take the help of the Lagrangian approach to predict the mixing both qualitatively as well as quantitatively. The results presented in this study, which we believe as the first to be obtained by the use of the variational calculus method, are in excellent qualitative agreement with the experimental observations (Hart 1971). For qualitative analysis of the mixing, we perform the Poincaré maps, whereas, for quantification, we estimate the variation of mixing index by the use of entropy of mixing.

For a fixed rotational speed of the channel $(\text{Re}_\omega)$, the Coriolis force, which also leads to the inviscid geostrophic core formation, becomes the sole function of the flow velocity. As a consequence, by increasing the magnitude of primary flow through strong electroosmotic pumping and a weak frictional drag, we achieve a considerable intensity of the geostrophic depression in the interior of the channel. The strong electroosmotic pumping in the present study is achievable for small DH parameters of EL $\kappa(=12)$ and higher PEL thickness $d(=0.3)$, whereas to realize the effect of a weak frictional drag on the underlying flow dynamics, we consider a small drag parameter of PEL $\alpha(=1)$. An increase in the intensity of the geostrophic depression in the core further results in a thin regime of the Ekman layer for the aforementioned cases [smaller $\kappa(=12)$, smaller $\alpha(=1)$, and higher $d(=0.3)$]. Such formation of thin Ekman layer for these cases [smaller $\kappa$, smaller $\alpha$, and higher $d$ ] further helps to achieve efficient mixing in the channel, even for the higher rotational speed of the channel. However, owing to the close competition between the strength of the corresponding electroosmotic force and its influencing regime, the change in the Ekman layer thickness is observed to be insignificant for the variation in $\kappa$.

Pertaining to the secondary flow analysis, we observe three specific vortex configurations: a standard double-vortex configuration for $\alpha = 20$ and $d = 0.05$; dumbbell-shaped vortices for $\kappa \sim [12-50]$, $d = 0.3$ and $\alpha = 1$; and transitional state between the first two vortex configurations for $d = 0.1$. Of these, the dumbbell-shaped vortices formed due to higher strength of Ekman pumping observed for strong electroosmotic effect in case of $\kappa = 12$, bring in more chaotic nature to the secondary flow, which in turn, establish an excellent mixing performance. Whereas, in contrast, by increasing the recirculation zone of the standard double-vortex configuration, the higher frictional drag for $\alpha = 20$ provides a good quality of mixing. Notably, the large recirculation zone of standard double-vortex configuration, which also appears in case of very small PEL thickness $d = 0.05$, plays an essential role to reduce the mixing length substantially. This is the key observation of this study. However, a notable compromise in the axial flow rate $(Q_x)$ obtained for higher drag parameter $\alpha = 20$ and very small PEL thickness $d = 0.05$ is suggestive of the restricted use of these cases ($\alpha = 20$ and $d = 0.05$) for the mixing enhancement. This inference underlines that the microchannel with built-in polymeric layers (supposedly provides higher drag in the PEL) or rigid microchannel (equivalence in very small PEL thickness) imbedded in CD platforms contradicts the prime objectives of achieving the maximum axial throughput and mixing performance simultaneously with equal priority. However, as shown in this analysis, this is not the case with variation in the DH parameter of EL.

The results of the present study also indicate that the tracer particles residing at the geostrophic core impede the process of mixing. The primary flow velocity at the geostrophic core, which substantially dominates its effect by imparting translational motion to this stream of particles, is the key cause for this retardation of the mixing process. More interestingly, it is also observed that the particles trapped inside the lobes of dumbbell-shaped vortices never interact with the particles of species co-injected at the inlet. For both the cases mentioned above, the information entropy at the geostrophic core and the lobes of the dumbbell-shaped vortices remain almost unaltered and minimum over the total length of the channel, thus augmenting the mixing length unreasonably. This aspect remains a challenge for further studies in this research paradigm.

With concern to this study, it seems reasonable to conjecture that the influence of the convective inertia on the flow dynamics and its consequences on the underlying mixing phenomenon is expected to bring significant weightage to the studied problem. For a sufficiently high Reynolds number, the standard double-vortex configuration or dumbbell-shaped vortices bifurcate into a non-trivial four-vortex configuration, thus expected to affect the mixing performance. In such cases, the elevated roll-cell instability is another fundamental concern (Kheshgi & Scriven 1985; Speziale 1982) to be accounted for on the underlying analysis. In a similar line, the effect of change in conformational entropy on the rotational soft microfluidics, primarily attributable to intra-molecular interaction inside the polyelectrolyte layer or an interaction between the polyelectrolyte and foreign bio-fluidic or chemical molecule is not thoroughly understood. In addition to that, the experimental investigation, which would mimic the present study, is worth pursuing in the future endeavour essentially for potential improvements of the rotating microfluidic devices/systems, largely used in biomedical/biochemical applications.

**Acknowledgements**

Authors wish to acknowledge High-performance computation (HPC) facility – *Param Ishan*, Computer and communication center of Indian Institute of Technology Guwahati. PKM acknowledges the financial support provided by the SERB (DST), India, Project No. ECR/2016/000702/ES. Authors also wish to thank Dr. Sweta Tiwari (Department of Mathematics, Indian Institute of Technology Guwahati) and Dr. P Kaushik (Department of Mechanical engineering, National Institute of Technology Tiruchirappalli) for their significant inputs and valuable suggestions.

Chen, G., & Das, S. (2017). Massively Enhanced Electroosmotic Transport in Nanochannels Grafted with End-Charged Polyelectrolyte Brushes. *The Journal of Physical Chemistry B*, **121**(14), 3130–3141.

Das, S., Banik, M., Chen, G., Sinha, S., & Mukherjee, R. (2015). Polyelectrolyte brushes: theory, modelling, synthesis and applications. *Soft Matter*, **11**(44), 8550–8583.

Ducrée, J., Haeberle, S., Brenner, T., Glatzel, T., & Zengerle, R. (2006). Patterning of flow and mixing in rotating radial microchannels. *Microfluidics and Nanofluidics*, **2**(2), 97–105.

Duffy, D. C., Gillis, H. L., Lin, J., Sheppard, N. F., & Kellogg, G. J. (1999). Microfabricated Centrifugal Microfluidic Systems: Characterization and Multiple Enzymatic Assays. *Analytical Chemistry*, **71**(20), 4669–4678.

Gaikwad, H. S., Kumar, G., & Mondal, P. K. (2020). Efficient electroosmotic mixing in a narrow-fluidic channel: the role of a patterned soft layer. *Soft Matter*, **16**(27), 6304–6316.

Gaikwad, H. S., Mondal, P. K., & Wongwises, S. (2018). Softness Induced Enhancement in Net Throughput of Non-Linear Bio-Fluids in Nanofluidic Channel under EDL Phenomenon. *Scientific Reports*, **8**(1), 1–16.

Hart, J. E. (1971). Instability and secondary motion in a rotating channel flow. *Journal of Fluid Mechanics*, **45**(02), 341.

Johnston, J. P., Halleent, R. M., & Lezius, D. K. (1972). Effects of spanwise rotation on the structure of two-dimensional fully developed turbulent channel flow. *Journal of Fluid Mechanics*, **56**(03), 533.

Kaushik, P., Abhimanyu, P., Mondal, P. K., & Chakraborty, S. (2017a). Confinement effects on the rotational microflows of a viscoelastic fluid under electrical double layer phenomenon. *Journal of Non-Newtonian Fluid Mechanics*, **244**(2), 123–137.

Kaushik, P., & Chakraborty, S. (2017). Startup electroosmotic flow of a viscoelastic fluid characterized by Oldroyd-B model in a rectangular microchannel with symmetric and asymmetric wall zeta potentials. *Journal of Non-Newtonian Fluid Mechanics*, **247**, 41–52.

Kaushik, P., Mondal, P. K., & Chakraborty, S. (2017b). Rotational electrohydrodynamics of a non-Newtonian fluid under electrical double-layer phenomenon: the role of lateral confinement. *Microfluidics and Nanofluidics*, **21**(7), 1–16.

Kaushik, P., Mondal, P. K., Kundu, P. K., & Wongwises, S. (2019). Rotating electroosmotic flow through a polyelectrolyte-grafted microchannel: An analytical solution. *Physics of Fluids*, **31**(2), 022009.

Keh, H. J., & Ding, J. M. (2003). Electrokinetic flow in a capillary with a charge-regulating surface polymer layer. *Journal of Colloid and Interface Science*, **263**(2), 645–660.

Kheshgi, H. S., & Scriven, L. E. (1985). Viscous flow through a rotating square channel. *Physics of Fluids*, **28**(10), 2968.

Lezius, D. K., & Johnston, J. P. (1976). Roll-cell instabilities in rotating laminar and trubulent channel flows. *Journal of Fluid Mechanics*, **77**(1), 153–174.

Masliyah, J. H. (1980). On laminar flow in curved semicircular ducts. *Journal of Fluid Mechanics*, **99**(3), 469–479.

Masliyah, J. H., & Bhattacharjee, S. (2006). *Electrokinetic and Colloid Transport Phenomena*, Wiley.

Nandakumar, K., Raszillier, H., & Durst, F. (1991). Flow through rotating rectangular ducts. *Physics of Fluids A: Fluid Dynamics*, **3**(5), 770–781.

Ng, C. O., & Qi, C. (2015). Electro-osmotic flow in a rotating rectangular microchannel. *Proceedings of the Royal Society A: Mathematical, Physical and Engineering Sciences*,46